\documentclass{article}

\usepackage{arxiv}

\usepackage[utf8]{inputenc} % allow utf-8 input
\usepackage[T1]{fontenc}    % use 8-bit T1 fonts
\usepackage{hyperref}       % hyperlinks
\usepackage{url}            % simple URL typesetting
\usepackage{booktabs}       % professional-quality tables
\usepackage{amsfonts}       % blackboard math symbols
\usepackage{nicefrac}       % compact symbols for 1/2, etc.
\usepackage{microtype}      % microtypography
\usepackage{lipsum}
\usepackage{cite}
\usepackage{amsmath,amssymb,amsfonts}
\usepackage{algorithmic}
\usepackage{graphicx}
\usepackage{textcomp}
\usepackage{xcolor}
\usepackage{bm}
\usepackage{url}
\usepackage{subfigure}
\usepackage{placeins}
\usepackage{epstopdf}
\usepackage{dsfont}
\usepackage{gensymb}
\usepackage{caption}
\usepackage{color,soul}

\newcommand{\E}{\operatorname{E}}

\newcommand{\R}{\mathds{R}}
\newcommand{\D}{\mathcal{D}}

\title{Modeling the Temporal Population Distribution of {\it Ae. aegypti} Mosquito using Big Earth Observation Data}

\author{
  Oladimeji Mudele\\
Telecommunications and Remote Sensing Laboratory, \\
Department of Electrical, Computer\\
and Biomedical Engineering,\\
University of Pavia, Italy\\
  \texttt{oladimeji.mudele@universitadipavia.it} \\
   \And
 F\'{a}bio M. Bayer \\
Department of Statistics\\
and LACESM, \\
Federal University of Santa Maria, Brazil\\
  \texttt{bayer@ufsm.br} \\
     \And
Lucas Zanandrez \\
  Ecovec company, Belo Horizonte, \\
  Brazil\\
  \texttt{lucas@ecovec.com} \\
     \And
Alvaro E. Eiras \\
Laboratory of Technological Innovation \\
and Entrepreneurship in Vector Control \\
Department of Parasitology, \\
Federal University of Minas Gerais, Brazil\\
  \texttt{alvaro@icb.ufmg.br} \\
     \And
Paolo Gamba\\
Telecommunications and Remote Sensing Laboratory, \\
Department of Electrical, Computer\\
and Biomedical Engineering,\\
University of Pavia, Italy\\
  \texttt{paolo.gamba@unipv.it} \\
}

\begin{document}
\maketitle

\begin{abstract}
Over 50\% of the world population is at risk of mosquito-borne diseases. Female \textit{Ae. aegypti} mosquito species transmit Zika, Dengue, and Chikungunya. The spread of these diseases correlate positively with the vector population, and this population depends on biotic and abiotic environmental factors including temperature, vegetation condition, humidity and precipitation. To combat virus  outbreaks, information about vector population is required. To this aim, Earth observation (EO) data provide fast, efficient and economically viable means to estimate environmental features of interest. In this work, we present a temporal distribution model for adult female \textit{Ae. aegypti} mosquitoes based on the joint use of the Normalized Difference Vegetation Index, the Normalized Difference Water Index, the Land Surface Temperature (both at day and night time), along with the precipitation information, extracted from EO data. The model was applied separately to data obtained during three different vector control and field data collection condition regimes, and used to explain the differences in environmental variable contributions across these regimes. To this aim, a random forest (RF) regression technique and its nonlinear features importance ranking based on mean decrease impurity (MDI) were implemented. To prove the robustness of the proposed model, other machine learning techniques, including support vector regression, decision trees and k-nearest neighbor regression, as well as artificial neural networks, and statistical models such as the linear regression model and generalized linear model were also considered. Our results show that machine learning techniques perform better than linear statistical models for the task at hand, and RF performs best. By ranking the importance of all features based on MDI in RF and selecting the subset comprising the most informative ones, a more parsimonious but equally effective and explainable model can be obtained. Moreover, the results can be empirically interpreted for use in vector control activities. 
\end{abstract}

% keywords can be removed

\section{Introduction}
Studying urban ecosystems is a hot topic within various scientific clusters \cite{Linard2013, donnay2000remote, Bagan2012, QihaoGamba2018a, Weng2014}. 
The compounding effects of human activities through urbanization, carbon emission and other biodiversity changes are modifying urban ecosystems, 
thus creating a need for continuous assessment of the environment to ensure health and quality of life suitability for dwellers \cite{While2013}. 
With a repository covering over 40 years of data collected, Earth observation (EO) data from orbiting satellites present a gold mine for environmental change monitoring \cite{Markham2012}. 
Due to recent advances resulting in higher resolution sensors (spectral, spatial and temporal), freely accessible data, and  efficient processing algorithms, it is possible to extract static and dynamic phenomena happening on the Earth surface and use this information for various environmental characterization procedures \cite{patel2015multitemporal, jiang2017spatial, Marinoni2017, Turyahikayo2015}.

One major area currently experiencing rapid innovation due to EO data is landscape epidemiology \cite{OSTFELD2005}. 
In this field, environmental variables that are proxies to favorable conditions for the spread of disease causing vectors are extracted from EO data and used to model the distribution of these vectors \cite{Reisen2010, Young2013}. Major disease carrier vectors with spread and population depending on environmental conditions include mosquito species, ticks and rodents \cite{Rossi2018,Johnson2017, Fornace2014, Jamison2015}. Mapping and predicting vector and diseases spread through EO data can provide reliable information in outbreak early warning, hotspots location, and public health budget optimization \cite{rotela2017analytical, Porcasi2012}. To this aim, there have been seminal works considering how to include major environmental variables obtainable from EO data such as surface temperature, vegetation condition, air quality, precipitation rate, humidity conditions, night-time lights, land cover types, degree of urbanization and velocity of the urban sprawl, among others \cite{Messina2016, Doll2006, Angiuli2014, Tran2019}.

According to the data provided by the World Health Organization, over 50\% of the world population is exposed to the risk of mosquito-borne diseases \cite{worldhealthorganization_2016, Jamison2015}, the ones with the greatest clinical importance being Zika, Chikungunya, Dengue and Yellow fever virus diseases. 
These diseases are transmitted mainly by the female \textit{Ae. aegypti} mosquito species \cite{rotela2017analytical, Johnson2017}. While its male counterpart feeds only on fruit nectar, the female \textit{Ae. aegypti} bites humans to feed on blood \cite{Scott1997} and poses  a serious threat to humans by transmitting dangerous viruses.
Areas of the world majorly exposed to risks of arbovirosis carried by this mosquito species include Latin America, Central Africa, and South-East Asia, with major disease spread outbreaks already recorded in Brazil, Argentina, Colombia, and Venezuela. 
The \textit{Ae. aegypti} species is known to be adaptable to urban environments due to its inclination to being bred in artificial containers \cite{rotela2017analytical, Messina2016, Scavuzzo2018, German2018}. Spread of the causal diseases by this vector is known to correlate with the local adult vector population \cite{antonio2017}. 
Environmental conditions including precipitation, vegetation conditions, temperature, and humidity have been shown in previous works to significantly influence \textit{Ae. aegypti} mosquito development \cite{Scavuzzo2018, Messina2016, Liu2017}. 
After obtaining these environmental variables from satellite data, 
statistical and machine learning (ML) 
models  
of diseases outbreak 
can be used for 
vector population prediction \cite{ritchie1984, eiras2018new}. 

ML is a subdivision of artificial intelligence which deals with the implementation of algorithms to learn complex patterns from machine readable input data for classification and regression purposes \cite{Lary2016,Karpatne2018}. 
It is a collection of methods that provide multivariate, nonlinear and non-parametric regression or classification of data. 
ML has proven useful for a very large number of EO system applications such as crop disease detection \cite{Yao2009}, urban area classification \cite{Angiuli2014}, landscape epidemiology \cite{ Scavuzzo2018, rotela2017analytical, Messina2016}, and various other bio-geographical information extraction and regression analysis procedures \cite{Lary2016, Karpatne2018}. 

In landscape epidemiology, 
ML algorithms can be used to map relationships between proxy environmental conditions and ground collected vector population information for the purpose of prediction, classification and/or explanation of environmental condition effects on vector spread \cite{Messina2016, Scavuzzo2018}. 
Previous attempts in modeling the population of female \textit{Ae. aegypti} and other species of mosquitoes have employed ML algorithms including 
maximum entropy methods, 
%boosted regression trees, 
support vector regression (SVR), 
decision trees regression (DTR), 
k-nearest neighbor regression (KNN), 
artificial neural networks (ANN), 
multilayer perceptron (MLP), 
among others \cite{rotela2017analytical, Messina2016, Porcasi2012, Scavuzzo2018}. 

In spite of the works in this study domain,  there are still some gaps. One of such is in the selecting the most informative environmental features subset to obtain the model with best fit and least redundancy. In \cite{Scavuzzo2018}, KNN, SVR, and MLP were used along with classical statistical models in time series modeling of {\it Ae. aegypti} oviposition activity. To select the best model, feature significance estimated with  linear correlation coefficient was used. This metric, however, only considers pairwise linear relationship among candidate features as selection criterion, ignoring nonlinear associations. The same limitation also applies to \cite{German2018}.  

In this paper,  we present a methodology for modeling the temporal population distribution of female {\it Ae. aegypti} mosquito based on time series environmental variables obtained from freely accessible EO data products and field collected mosquito population data. 
To achieve our aim, a
random forest (RF) \cite{Breiman2001,Biau2016}
model is considered due to: 
(i) its robustness against unbalanced data 
with good performance in complex problems 
and 
(ii) its embedded mean decrease impurity (MDI) variable importance measure. 
MDI is used to rank and extract the most relevant environmental feature subset for modeling the vector population. It considers also nonlinear associations between candidate features. 
For benchmark purposes, 
RF is compared with other ML models already used in similar works 
namely: KNN, SVR, DTR, and MLP, 
and statistical regression models 
such as linear regression model (LM) 
and generalized linear model (GLM) \cite{McCullagh1989}. Finally, we considered the best model, fitted with the most informative features as ranked by MDI, and demonstrated using our data how the effects of the selected features on the vector population can be explained and made into actionable insights for field investigators.

\section{Study Area and EO variables}\label{study_area_EO}

Even though climatic and vegetation condition effects alone do not determine the population of \textit{Ae. aegypti} mosquitoes,  entomological and epidemiological studies have described the effects of temperature, humidity, precipitation, and vegetation condition  on the population and virus transmission factors of \textit{Ae. aegypti} \cite{Jansen2010}. 
In this study,  we model the population of mosquitoes as a function of selected remotely sensed environmental variables. 

\begin{figure}
\centering
\includegraphics[width=0.8\textwidth]{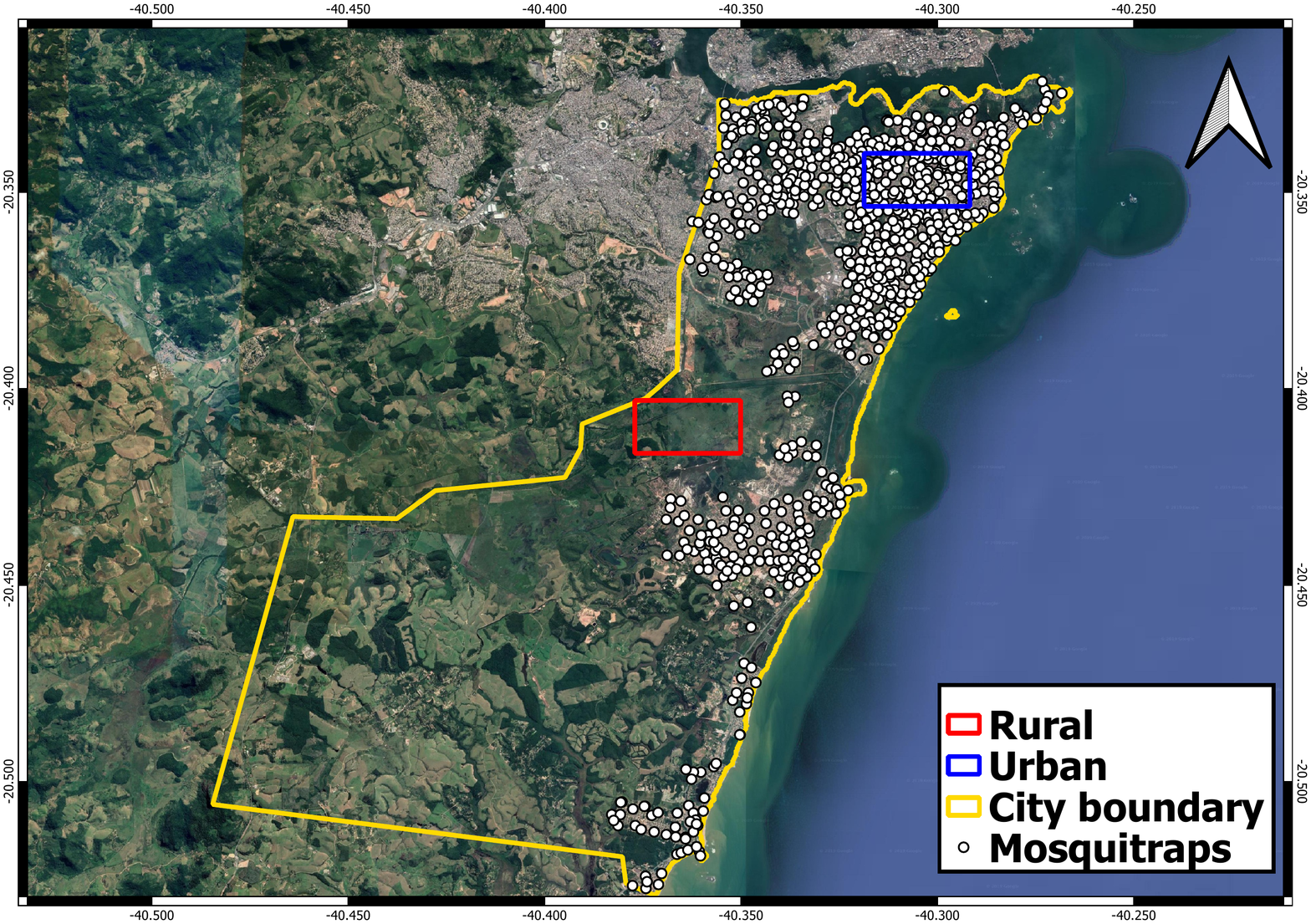}
\caption{Aerial view of Vila Velha municipality, data collection mosquito trap locations, and urban and rural surface zones selected to extract the environmental variables.}
\label{f:area}
\end{figure}

To this aim, we consider spatiotemporal data for the municipality of Vila Velha, located between latitudes $20\degree$ $19^\prime$ and $20\degree$ $32^\prime$ south and longitudes $40\degree$ $16^\prime$ and $40\degree$ $28^\prime$ west, on the coast of the Esp\'irito Santo State of Brazil.  This municipality covers a total area of $209,965 km^2$, with an estimated population of $479,664$ inhabitants \cite{Santos2017}.  Figure \ref{f:area} shows the location of the studied area. 

The control of \textit{Ae. aegypti} borne diseases is carried out in Brazil based on the National Dengue Control Program (PNCD) which has been in operation since 2002 \cite{costa2018national}. This program addresses important components including: epidemiological surveillance, vector control, environmental sanitation, among others \cite{araujo2015aedes}. Plugging into the epidemiological surveillance framework, Ecovec --- an expert real-time mosquito surveillance company based in Belo Horizonte, Brazil \cite{aboutecovec} --- began the implementation of the MI-Aedes\textsuperscript{\textregistered} program in Vila Velha in 2017 to improve the operation of PCND. The MI-Aedes\textsuperscript{\textregistered} program uses 791 geopositioned adult mosquito traps to obtain weekly \textit{Ae. aegypti} population data across the municipality  of Vila Velha. A single mosquito trap is placed at every point, with approximately 250 meters separation from the nearest surrounding traps to limit the effect of spatial autocorrelation. All the utilized traps are of the same type, specifically, the sticky-card oviposition trap, commercially named MosquiTrap\textsuperscript{\textregistered} \cite{eiras2009preliminary}. This type of trap is made of a black bucket containing water and a sticky-card to capture any mosquitoes that try to lay eggs on the trap's internal walls. It must be inspected weekly to change the water and collect mosquito population data. Also, a commercial mosquito lure (AtrAedes\textsuperscript{\textregistered}) was used to enhance attraction of females of Aedes aegypti to the trap. An aerial view of the used MosquiTraps\textsuperscript{\textregistered} location is shown in Figure \ref{f:area}.

Since the commencement of MI-Aedes\textsuperscript{\textregistered},  vector control actions in Vila Velha has achieved improvements at intervals by using the mosquito population data collected. For example, data collected in 2017 was used to improve control actions from the 1st week of 2018 by directing more control actions to places with higher adult mosquito population. One major advantage of the  MI-Aedes\textsuperscript{\textregistered} program is that it collects weekly data of adult vector population, not immature forms. This ensures that the data highly correlates with human diseases infection cases since only adult female mosquitoes are responsible for the transmission \cite{Melo2012}. 

For the purpose of this work, we used weekly female \textit{Ae. aegypti} MI-Aedes\textsuperscript{\textregistered} program data from week 15 in 2017 to week 34 in 2019 --- a total of 124 weeks. Data collection was carried out during weekly inspections conducted by  municipal field workers who are trained by Ecovec for trap inspections, mosquito species identification, and to use an online mobile system to send the data. The inspections occurred on the same day of the week, and same time of the day for each MosquiTrap\textsuperscript{\textregistered}. The collected and sent data were weekly checked and approved by a trained supervisor and a specialist from Ecovec. To maintain operational standards on the field during the collection tenure, an Ecovec specialist visited the field every six months to check  the conditions of the traps and to update staff training. 

Our analyses account for changes in data collection conditions such as the earlier reported data-driven improvement of control actions starting from the 1st week of 2018 and an update in the MI-Aedes\textsuperscript{\textregistered} platform starting from week 42 in 2018 which led to changes in identification number of the mosquito traps and slight changes in their geopositions.  In the cases of trap location changes, the traps were moved to new adjacent residences due to monitoring difficulties or because residents specifically requested the changes. All in all, trap relocations beyond a 40 meters radius were rare.

As a result of these discrepancies in data collection conditions, our data and consequent analyses were split in three time batches, presented in Table \ref{t:databatches}. Since previous works \cite{Scavuzzo2018, German2018} have shown that it is sufficient to work with weekly observation samples, we cleaned the \textit{Ae. aegypti} population data  to obtain female mosquito population per mosquito trap on a weekly basis. It is important to note two points here: firstly, we expect lower vector population and correlation with environment condition starting from the second Batch in our analyses due to improvements in control actions. Secondly, the three data Batches do not exactly span the same time and seasons of the year.

\begin{table}%[ht]
\renewcommand{\arraystretch}{1.3}
\centering
\caption{Mosquito population data Batches used in this study
}
\label{t:databatches}
\begin{tabular}{llll}
  \hline
Batch & Date range & Total  & Differentiating\\ 
 & &weeks &condition \\
  \hline
1 & 10/04/2017 - 31/12/2017 & 38 & - \\
\hline
2 & 02/01/2018 - 05/10/2018 & 40 & Vector control \\
 & & & improvement \\
 \hline
3 &  08/10/2018 - 23/08/2019 & 46 & Update to the  \\
 & & & MI-Aedes\textsuperscript{\textregistered} platform \\
\hline
\end{tabular}
\end{table}

\begin{figure}
\centering
\includegraphics[width=0.7\textwidth]{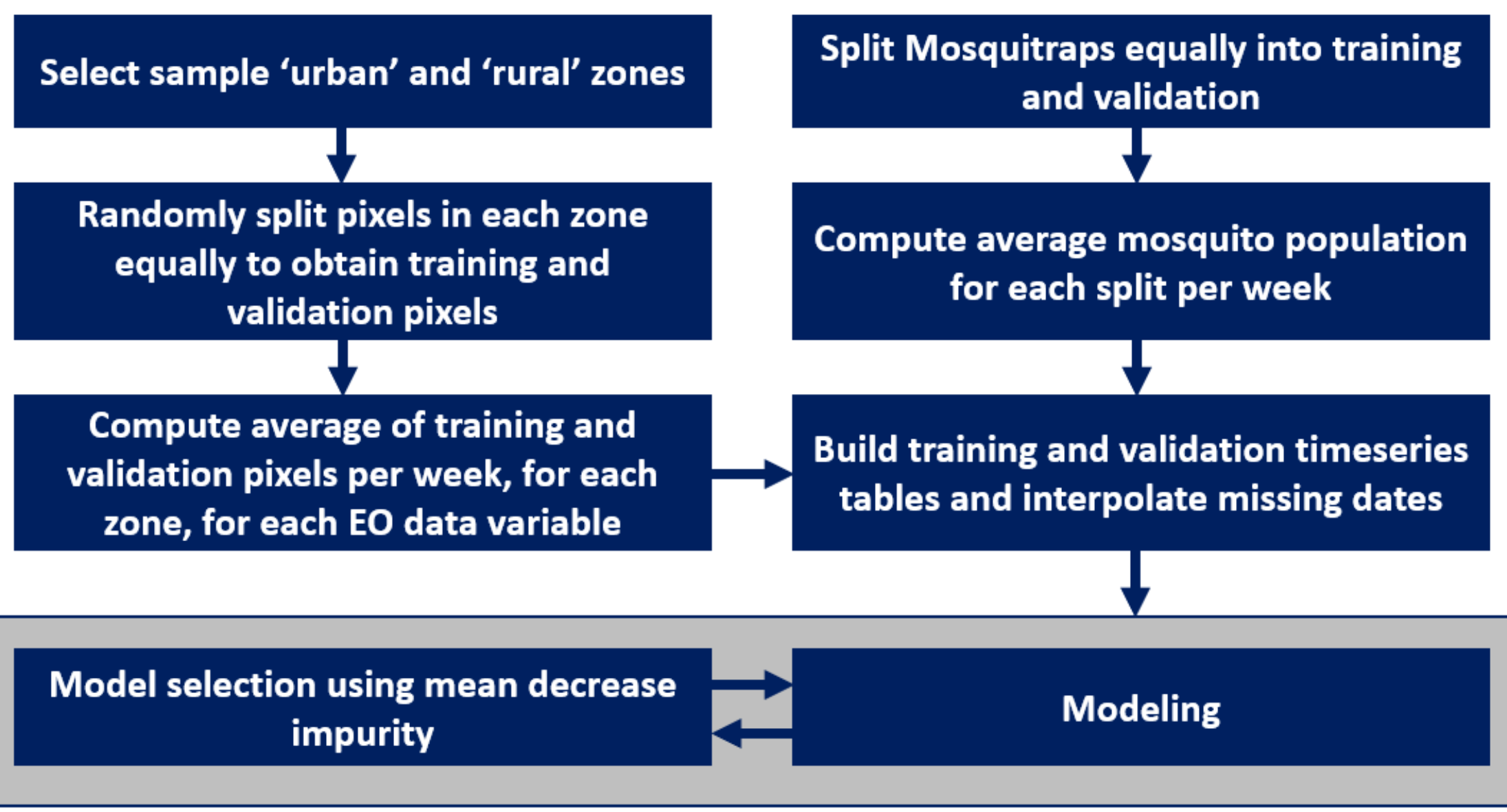}
\caption{Schematic of the study methodology.}
\label{f:processing_chain}
\end{figure}

In studies that associate the development of \textit{Ae. aegypti} with EO derived environmental factors, the Moderate Resolution Imaging Spectroradiometer (MODIS) satellite data products provided by the National Aeronautics and Space Administration (NASA) have been recently used \cite{Scavuzzo2018, German2018}. This is due to its free access, global availability and non-requirement of much further processing to obtain the needed environmental features \cite{Justice2002}.

For this work, data products providing information corresponding to surface temperature, precipitation, humidity, and vegetation conditions were collected for the weeks matching the vector population data.  For vegetation condition information, the Normalized Difference Vegetation Index (NDVI) has been obtained from the MODIS MOD13Q1 product -- a 16-day composite product with spatial resolution of 250 meters.  The near and mid-infrared bands of the same data product was used to compute the Normalized Difference Water Index (NDWI) using Gao's definition \cite{Gao1996}. This layer provides information about vegetation water content which is a proxy to humidity and surface moisture \cite{Scavuzzo2018}. In addition, estimates of the minimum and maximum surface temperatures have been obtained from the day and night-time land surface temperature (LST) bands of the 8-day composite MODIS MOD11A2 product at 1 km spatial resolution \cite{Wan2004}. Finally, the calibrated precipitation band of the Global Precipitation Measurement (GPM) mission data product was used to obtain precipitation information \cite{skofronick2017global}, although in a very coarse way, because this data is available daily at $0.1\deg$ spatial resolution ($\approx$ 11 km). All data were downloaded using the \texttt{export} object of the \texttt{JavaScript} application programming interface of google earth engine (GEE). This object has resampling and reprojection methods abstracted into it for easy co-registration of data with different properties. For each download task, we obtained the output data at a resampled spatial resolution of 250 meters because this is the minimum distance separating neighboring mosquito traps. We used the \texttt{scale} parameter of the \texttt{export} object to set this spatial resolution value. We set the common coordinate reference system projection for all downloaded data as \texttt{EPSG:4326} using the \texttt{crs} parameter.

As previously implemented in \cite{Scavuzzo2018, German2018} and \cite{Estallo2008}, to find the  relationship between the measured number of mosquitoes and the model output exploiting EO data, we defined two buffer areas of size 18 $km^2$ each from which distinct temporal characterization of the considered EO variables are obtained within our study area. One of the zones is in the densely urbanized part of the city (labeled as "urban"), and the other in the peri-urban zone (labeled as "rural") areas. These two areas are shown in Figure~\ref{f:area}. 

As in typical supervised ML applications, 
pixels within each selected zone were randomly and equally
split into the training and validation sets. 
The average for training and validation pixels were computed for each image in the stack of the EO data set representing individual environmental variables considered. 
To obtain features for the mosquito data, we aggregated the training and validation data for each week from all the mosquito traps in the whole study area, not only from the buffer zones as in the case of the EO covariates.   We randomly split all the mosquito traps equally into training and validation. We compute the means for each split per week to obtain the response time-series training and validation data. We used a fifth order spline interpolation to obtain values for missing dates.
All environmental variables are considered with up to 2 weeks lag to account for non-synchronous effects resulting from vector development life cycle \cite{cheong2013assessing}. 
As in \cite{Scavuzzo2018}, all the variables were standardized (z-score). 
This was done because variable rescaling is a good practice, especially for training a neural network \cite{Bishop1996}. A high-level schematic of the whole methodology of our study as described in this section is presented in Figure \ref{f:processing_chain}.

\section{Modeling Procedure}\label{s:models}

To model the 
weekly mean number of mosquitoes per mosquito trap ($Y$),
we consider fitting a RF model. 
For prediction benchmark purpose, ANN, SVR, KNN, DTR, LM, and GLM fitting models were also considered. Finally, a more parsimonious RF model, labeled $\text{RF*}$, considering only the  most informative climate variables  obtained by using the MDI to rank all predictor covariates was also implemented and compared with the other ones.
For prediction performance comparison, the usual quantitative measures  between the observed data ($y$) and the predicted values ($\hat{y}$) were considered, namely:  the linear correlation coefficient (R),  the root mean square error (RMSE),  and the mean absolute percentage error (MAPE) \cite{Sammut2017}. We use R and MAPE to measure relative qualities, and RMSE to measure absolute fit of our models.  Background details and our implementation (function and model parameters) of RF are presented in Section \ref{RF}.  We present the same for the other models in Section \ref{s:other}.  We implemented all our {modeling} computation with {\tt R} programming \cite{Rlanguage}. 

\subsection{Random Forest Regression} \label{RF}
Random forests (RF) \cite{Breiman2001} are one of the most effective computationally intensive procedures  to improve on unstable estimates, especially when it is difficult to find a good model due to problem complexity \cite{Biau2016}. It is a predictor consisting of a collection of several randomized regression trees \cite{Biau2016}. 
Let $\mathbf{X} \in \chi \subset \R^p$ an input vector related to $p$ features
and 
$Y \in \R$ the response random variable, 
the objective is to estimate the regression function
$m(\mathbf{x})=\E(Y|\mathbf{X}=\mathbf{x})$. 
Given a training sample 
$\D_n = ((\mathbf{X}_1,Y_1),\ldots,(\mathbf{X}_n,Y_n))$ 
of independent random variables distributed as the independent prototype pair $(\mathbf{X},Y)$, 
the goal is to use the data set $\D_n$ to construct an estimate 
$m_n: \chi \rightarrow \R$ 
of the function $m$. 
To this aim, 
the random forest consists of a collection of $M$ randomized regression trees. 
For the $j$th tree in the family, with $j=1,\ldots,M$, 
the predicted value at each $j$ is denoted by 
$m_n( \mathbf{x}; \Omega_j,\D_n)$, 
where $\Omega_1,\ldots,\Omega_M$ 
are random variables independent of $\D_n$. 
The variables 
$\Omega_j$ 
are used to resample the training set prior to the growing of individual trees 
and to
select the directions for splitting. 
Then, 
the trees are combined to form the forest estimate given by:
\begin{align}
\hat{y}=m_{M,n}( \mathbf{x}; \Omega_1,\ldots,\Omega_M,\D_n)
=
\frac{1}{M}\sum_{j=1}^M m_n( \mathbf{x}; \Omega_j,\D_n).
\end{align}

In this procedure, we have also to set other tuning parameters, 
i.e.,
the number of possible directions for splitting ($m_s$) at each node of each tree, 
and 
the lowest number of examples ($m_e$) in each cell to cause a split. 
In this work 
we set initially $M=500$, 
$m_s=\lceil p/3 \rceil$ 
and 
$m_e=5$. 

One important characteristic 
of the random forest 
is that 
it can be used to rank 
the importance of the input variables (i.e., the features we extract from EO data) 
over the pattern variability of the target variable $Y$. 
In this work,
we considered the 
MDI
calculated based on 
the reduction in sum of prediction squared errors  
averaged over all trees  
whenever a variable is chosen to split \cite{louppe2013understanding}.

\subsection{Other Predictive Models}\label{s:other}
As mentioned above, for comparison purposes a number of other regression models have been implemented and their results compared with the ones by the proposed RF procedure. Specifically, they are:
\begin{itemize}

\item the Linear Regression Model (LM): the most used predictive model in different fields. 
With respect to the topic of this paper, the LM has been used in similar works, \cite{German2018} and \cite{Scavuzzo2018},  to model the \textit{Ae. aegypti} oviposition activity in a northern Argentine city. In this work, LM is implemented using the {\tt lm()} function in {\tt R},  and an ordinary least square estimator is considered to estimate the regression parameters. 

\item The Generalized Linear Model (GLM): a class of regression models able to model response variables in the exponential family of distributions \cite{McCullagh1989}.  Since the weekly mean value of mosquitoes population ($y$) is continuous and always in the space of positive real numbers ($y \in \R^+$),  the Gaussian assumption considered for inference in the LM can lead to poor results and predictions. In GLM, instead, the gamma distribution is considered to model $y$, together with the log link function in the GLM regression structure. 

\item Artificial Neural Networks (ANN): one of the most used ML methods for geoscience problems. More recently, deep learning methods -- corresponding to ANN architectures with several hidden layers \cite{Karpatne2018} -- became widely applied in many EO data processing problems \cite{Lary2016, Zhang2016}. The ANN is a nonlinear data modeling technique used to model complex relationships between sets of input and output variables \cite{Lary2010}. In this work, a multilayer perceptron with three layers, each with three neurons, is used. Each of the neuron is activated with a  logistic activation function, and the resilient backpropagation with weight backtracking algorithm  \cite{Riedmiller1994} was considered for training the neural network. 

\item Support Vector Regression (SVR): an approach based on support vectors with radial basis function kernel with tuning parameter $\gamma$ set as 1/(number of features), tuned via the the epsilon-regression method. 

\item k-Nearest Neighbor Regression (KNN): a regression based on the 4-nearest neighbors according to Euclidean distance and uniform weights for local interpolation. 

\item Decision Trees Regression (DTR): a regression model in the form of a tree structure \cite{Breiman1984}. It is a simple but powerful prediction method \cite{Krzywinski2017}.
\end{itemize}

\section{Results and Discussion} 
\begin{figure}
\centering
\includegraphics[width=0.5\textwidth]{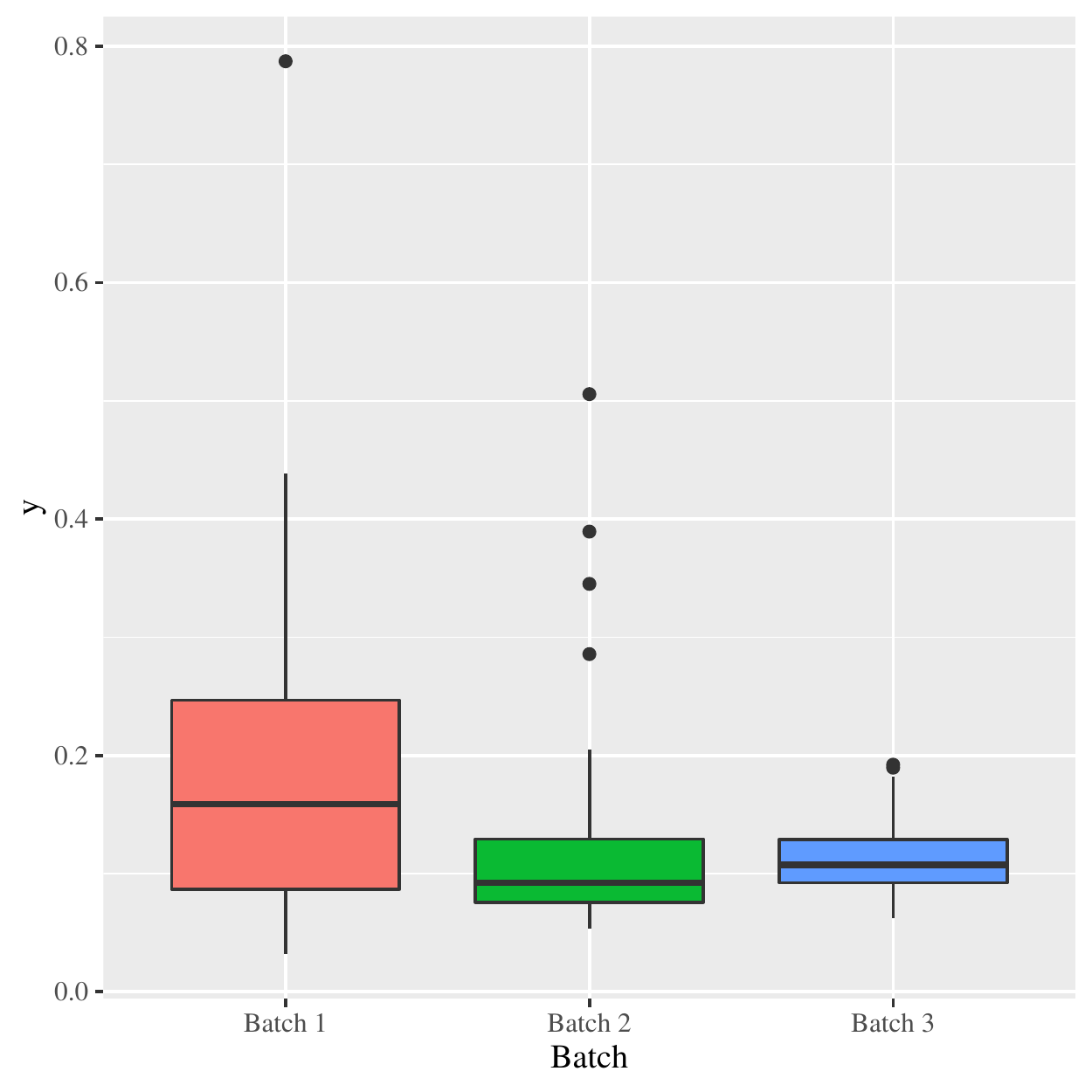}
\caption{Boxplot comparing the distribution of the number of female mosquitoes in the three considered periods.}
\label{f:boxplot}
\end{figure}

All the EO data variables used in this work were considered both in an urban surface zone (U)  and in a peri-urban, or rural zone (R).  We thus have the following variables: NDVI-U, NDWI-U, TempD-U, TempN-U, and Prec-U,  as well as NDVI-R, NDWI-R, TempD-R, TempN-R, and Prec-R, where TempD is the daytime temperature and TempN is the night-time temperature.   Table \ref{t:descriptive} reports the mean, median, standard deviation (SD), minimum (Min) and  maximum (Max)  of the EO data variables together with the  female mosquito population ($y$).  These measures were computed for the three Batches of time considered in our study (cf. Table~\ref{t:databatches}) in a separate way, to evaluate the influence of the vector control program improvement implemented in 2018 and 2019, affecting Batches 2 and 3.  We observed that the average value of the number of female mosquitoes in Batch 1 is 0.1910,  while the values in Batches 2 and 3 are 0.1267 and 0.1134, respectively. This represents about $40\%$ decrease in the mean number of mosquitoes from Batch 1 compared to subsequent Batches considered and highlights the efficacy of the data-driven vector control program improvement which has been implemented.
%Even better, considering the median, nearly a $50\%$ decrease is observed. 
Furthermore, the nonparametric Kruskal-Wallis test \cite{Hollander2013} 
rejects the null hypothesis of three populations being equally distributed ($p$-value = 0.0092), confirming that the decrease in the mean number of mosquitoes from Batches 1 to 2 and 3 is statistically significant. 
In addition to the Kruskal-Wallis test, 
the Nemenyi test shows that the population from Batch 1 differs from Batches 2 and 3, 
but population distributions of mosquitoes of Batches 2 and 3 do not differ statistically. 
Also, 
we can note that the standard deviation of the number of mosquitoes decreases from 
0.1446 in Batch 1 to 0.0956 and 0.0315 in Batches 2 and 3, respectively. 
The boxplot in Figure~\ref{f:boxplot} shows the differences among the population of female mosquitoes in the different Batches. 
Particularly, 
we can see that the maximum value in Batch 3 is four times less than the maximum number in Batch 1.

The environmental characteristics, as presented in Table \ref{t:descriptive}, slightly differ across the three considered batch periods due to differences in observation time and seasons. Also, it can be seen that there was more precipitation in 2017 (Batch 1) than 2018 and 2019 (Batches 2 and 3).  Regarding the urban and rural zones, the average land surface temperature in urban zone, TempD-U and TempN-U, respectively, is, as expected, higher and more variable than in the rural zone. Moreover, as equally expected, the opposite is the case for both NDVI and NDWI.

\begin{table}%[ht]
\footnotesize
\centering
\caption{Descriptive measures of the EO-based environmental variables of interest, computed separately for each Batch
}
\label{t:descriptive}
\begin{tabular}{rrrrrr}
  \hline
Variable & Mean & Median & SD & Min & Max \\ 
  \hline
\multicolumn{6}{c}{Batch 1 (2017)} \\ 
  \hline
  $y$ & 0.1910 & 0.1587 & 0.1446 & 0.0319 & 0.7873 \\ 
  NDVI-U & 0.2269 & 0.2260 & 0.0261 & 0.1702 & 0.2841 \\ 
  NDVI-R & 0.7280 & 0.7395 & 0.0448 & 0.5826 & 0.7983 \\ 
  NDWI-U & 0.0024 & 0.0035 & 0.0350 & -0.0994 & 0.0748 \\ 
  NDWI-R & 0.5407 & 0.5492 & 0.0473 & 0.3952 & 0.6640 \\ 
  TempD-U & 31.4166 & 31.0544 & 4.1390 & 22.5969 & 39.3821 \\ 
  TempD-R & 27.7208 & 27.7761 & 2.9338 & 20.5000 & 33.8600 \\ 
  TempN-U & 20.0180 & 20.6767 & 2.4379 & 13.4942 & 23.5909 \\ 
  TempN-R & 19.0909 & 19.3264 & 1.8757 & 13.6278 & 23.1461 \\ 
  Prec-U & 2.9118 & 0.2000 & 7.5662 & 0.0000 & 43.2500 \\ 
  Prec-R & 2.9039 & 0.5375 & 6.7846 & 0.0000 & 37.1750 \\ 
  \hline
\multicolumn{6}{c}{Batch 2 (2018)} \\ 
  \hline  
  $y$ & 0.1267 & 0.0923 & 0.0956 & 0.0531 & 0.5057 \\ 
  NDVI-U & 0.2500 & 0.2474 & 0.0179 & 0.2086 & 0.2996 \\ 
  NDVI-R & 0.7475 & 0.7554 & 0.0261 & 0.6802 & 0.7922 \\ 
  NDWI-U & 0.0030 & -0.0021 & 0.0365 & -0.0591 & 0.1065 \\ 
  NDWI-R & 0.5923 & 0.5965 & 0.0518 & 0.4762 & 0.6925 \\ 
  TempD-U & 31.8615 & 31.5774 & 3.4781 & 26.4761 & 38.8394 \\ 
  TempD-R & 27.5138 & 26.9434 & 2.6868 & 23.7850 & 33.8161 \\ 
  TempN-U & 21.2936 & 21.0190 & 2.2088 & 16.3886 & 25.3319 \\ 
  TempN-R & 20.1959 & 19.6536 & 2.0855 & 16.5067 & 23.7806 \\ 
  Prec-U & 2.4707 & 0.0654 & 4.9280 & 0.0000 & 23.0803 \\ 
  Prec-R & 1.7527 & 0.0990 & 3.4678 & 0.0000 & 16.0950 \\ 
  \hline
\multicolumn{6}{c}{Batch 3 (2018-2019)} \\ 
  \hline  
  $y$ & 0.1134 & 0.1075 & 0.0315 & 0.0620 & 0.1922 \\ 
  NDVI-U & 0.2557 & 0.2551 & 0.0188 & 0.2117 & 0.3075 \\ 
  NDVI-R & 0.7210 & 0.7265 & 0.0331 & 0.5843 & 0.7829 \\ 
  NDWI-U & -0.0137 & -0.0200 & 0.0324 & -0.0762 & 0.0661 \\ 
  NDWI-R & 0.5350 & 0.5450 & 0.0538 & 0.4031 & 0.6272 \\ 
  TempD-U & 33.6604 & 33.8508 & 4.3575 & 18.6561 & 39.0947 \\ 
  TempD-R & 28.9434 & 29.4046 & 2.3435 & 24.7489 & 33.1411 \\ 
  TempN-U & 22.0600 & 22.4421 & 2.1145 & 17.3431 & 25.2092 \\ 
  TempN-R & 20.6418 & 21.1400 & 2.6758 & 13.2544 & 24.1517 \\ 
  Prec-U & 1.6132 & 0.1596 & 2.8811 & 0.0000 & 11.3604 \\ 
  Prec-R & 1.3545 & 0.0419 & 2.4574 & 0.0000 & 9.5575 \\ 
\hline
\end{tabular}
\end{table}

The linear correlation between each environmental variable, including lagged ones, and the female mosquito population, $y$, is presented in Table \ref{t:cor}. This table shows, first of all, that none of the considered covariates show very strong linear correlation (above 0.7) with $y$. However, in all Batches, the non-lagged daytime temperature variables show the highest correlation with $y$: TempD-R and TempD-U --- in that order --- in both Batches 1 and 2, and only the former in Batch 3. In general, more EO variables show higher correlation with the vector population in Batch 2, with five non-lagged variables showing correlation above 0.4, compared to the other Batches.  %It is clear that the linear correlation between the mosquito population is larger, for specific variables, only for peculiar values of the time lag lagged variables.

In addition, we can note that the lagged NDWI-R and NDVI-R and the precipitation variables show greater correlation with $y$ than with non-lagged ones. 
%For instance, in Batch 3  the correlation between $y$ and NDWI-R (lag zero) is $0.0380$ 
%while the correlation with lag two is $0.1480$.  
This observation can be explained by the aquatic cycle of \textit{Ae. aegypti} from egg to adult being approximately 7-9 days\cite{cheong2013assessing}, thus making the transition dependent on non-synchronous environmental effects, as also highlighted in \cite{chen2010lagged}.
We observe that the linear relationships between the targets and the predictors could be insufficient to describe the temporal variance in the mosquito population,  justifying the need for non-linear approaches such  as the considered ML methods.

\begin{table}%[ht]
\footnotesize
\centering
\caption{Correlation matrix between the average number of mosquitoes ($y$) and each covariate among the pool of EO-based variables in Table~\ref{t:descriptive}
}
\label{t:cor}
\begin{tabular}{rrrr}
\hline
Covariate	&	Lag 0	&	Lag 1	&	Lag 2	\\
	\hline
\multicolumn{4}{c}{Batch 1} \\ 
\hline	  
NDVI-U & -0.3097 & -0.0926 & -0.1127 \\ 
  NDVI-R & 0.0030 & 0.0958 & -0.0142 \\ 
  NDWI-U & -0.3989 & -0.2592 & -0.1480 \\ 
  NDWI-R & 0.0624 & 0.1566 & 0.1342 \\ 
  TempD-U & 0.4046 & 0.3086 & 0.2236 \\ 
  TempD-R & 0.4157 & 0.3327 & 0.1532 \\ 
  TempN-U & -0.2959 & 0.0895 & 0.1403 \\ 
  TempN-R & 0.1105 & 0.1701 & 0.0328 \\ 
  Prec-U & 0.0084 & -0.0349 & -0.0553 \\ 
  Prec-R & 0.0261 & -0.0368 & -0.0335 \\ 
\hline
\multicolumn{4}{c}{Batch 2} \\ 
\hline	
NDVI-U & -0.1401 & -0.2117 & -0.3098 \\ 
  NDVI-R & -0.5025 & -0.5852 & -0.4079 \\ 
  NDWI-U & -0.2454 & -0.2772 & -0.2118 \\ 
  NDWI-R & -0.2702 & -0.3954 & -0.3392 \\ 
  TempD-U & 0.5810 & 0.4700 & 0.3565 \\ 
  TempD-R & 0.6418 & 0.4962 & 0.4988 \\ 
  TempN-U & 0.4835 & 0.4150 & 0.3578 \\ 
  TempN-R & 0.4323 & 0.3584 & 0.4254 \\ 
  Prec-U & -0.0185 & -0.1644 & 0.0360 \\ 
  Prec-R & 0.0053 & -0.1595 & -0.0232 \\
  \hline
\multicolumn{4}{c}{Batch 3} \\ 
\hline	
NDVI-U & -0.1527 & -0.0546 & 0.0913 \\ 
  NDVI-R & 0.0439 & 0.0807 & 0.3296 \\ 
  NDWI-U & 0.1716 & 0.0320 & -0.0873 \\ 
  NDWI-R & 0.0380 & 0.1005 & 0.1480 \\ 
  TempD-U & -0.1082 & -0.0414 & -0.0750 \\ 
  TempD-R & -0.2703 & -0.1654 & -0.2530 \\ 
  TempN-U & -0.1004 & -0.0566 & -0.0871 \\ 
  TempN-R & -0.1352 & -0.1698 & -0.0909 \\ 
  Prec-U & -0.0969 & -0.1376 & 0.0578 \\ 
  Prec-R & -0.0384 & -0.1142 & 0.0589 \\
\hline	
\end{tabular}
\end{table}

Considering the results for the validation dataset,  the quantitative measures of the prediction obtained from each considered model  are shown in Table \ref{t:quality-measures}  for all Batches.
In Batch 1, the measures  show that GLM produces the worst results in terms of R and RMSE,  while LM the worst in terms of MAPE. The best performances are reached in R by RF, in RMSE by $\text{RF}^*$, and in MAPE by ANN.  The outstanding performance of RF, and its more parsimonious variant, $\text{RF}^*$,  is highlighted by the correlation measure of 0.90 and 0.86, and RMSE of 0.0369 and 0.0396, respectively.  Also, in Batch 2, we obtain the best result in terms of RMSE with $\text{RF}^*$, followed by the fully fitted RF. In this Batch, however, best results  considering R and MAPE are obtained using KNN. In Batch 3, $\text{RF}^*$ produced the best model as measured by R, RMSE and MAPE. Considering RMSE, $\text{RF}^*$ produces equal or better quality in all Batches. In general, from the whole results, though the statistical models in certain cases produced sufficient or comparable measured qualities,  there are much more cases of better performances by machine learning models. For example, LM produces comparable quality with respect to RF in all Batches of our analyses if we look only at the explained variance which is measured by R. When we consider the RMSE, however, we see that that RF performs better than LM in terms of absolute fit. LM's high correlation measure of 0.8062 and 0.8535 in Batches 1 and 2 respectively are as a result of better correlation between more of their EO covariates and $y$ compared to what is obtained in Batch 3. The much lower R measure for the all models on Batch 3 data could be as a result of time cumulative effect of data driven improvement in the vector control actions which led to less variation in the mosquito population due to environmental changes. This same cause may be used to explain the lower correlation among more EO covariates and $y$ in the Batch 3 with respect to Batches 1 and 2, as earlier presented in Table \ref{t:cor}.

\begin{table}%[ht]
\footnotesize
\centering
\caption{Quality measures of predictions in the validation dataset}
\label{t:quality-measures}
\begin{tabular}{l|rrr}
  \hline
Model & R & RMSE & MAPE \\ 
  \hline
  & \multicolumn{3}{c}{Batch 1} \\
  \hline
  GLM & 0.6274 & 0.0722 & 60.6487 \\ 
  LM & 0.8062 & 0.0480 & 75.0815 \\ 
  ANN & 0.8420 & 0.0447 & 34.1122 \\ 
  SVR & 0.8270 & 0.0471 & 36.1444 \\ 
  KNN & 0.7456 & 0.0478 & 64.1194 \\ 
  DTR & 0.6334 & 0.0547 & 67.9013 \\ 
  RF & 0.9066 & 0.0369 & 56.9834 \\ 
  $\text{RF}^*$ (8 features)  & 0.8618 & 0.0396 & 61.8109 \\ 
  \hline
  & \multicolumn{3}{c}{Batch 2} \\
   \hline
  GLM & 0.6366 & 0.0490 & 43.0603 \\ 
  LM & 0.8535 & 0.0309 & 44.3038 \\ 
  ANN & 0.7202 & 0.0360 & 44.3968 \\ 
  SVR & 0.8651 & 0.0252 & 30.9242 \\ 
  KNN & 0.8839 & 0.0223 & 30.7137 \\ 
  DTR & 0.7925 & 0.0281 & 32.5429 \\ 
  RF & 0.8530 & 0.0255 & 35.4397 \\ 
  $\text{RF}^*$ (8 features) & 0.8594 & 0.0247 & 33.4977 \\ 
  \hline
  & \multicolumn{3}{c}{Batch 3} \\
   \hline
  GLM & 0.5707 & 0.0180 & 23.8297 \\ 
  LM & 0.5697 & 0.0168 & 23.3971 \\ 
  ANN & 0.4048 & 0.0192 & 24.3962 \\ 
  SVR & 0.4096 & 0.0174 & 23.8113 \\ 
  KNN & 0.3662 & 0.0180 & 27.1013 \\ 
  DTR & 0.3466 & 0.0179 & 25.7221 \\  
  RF & 0.5981 & 0.0156 & 21.4197 \\ 
  $\text{RF}^*$ (8 features) & 0.5885 & 0.0156 & 21.5879 \\ 
  \hline
\end{tabular}
\end{table}

\begin{figure}
\centering
\subfigure[Batch 1]{\label{f:scatter1}  
\includegraphics[width=70mm]{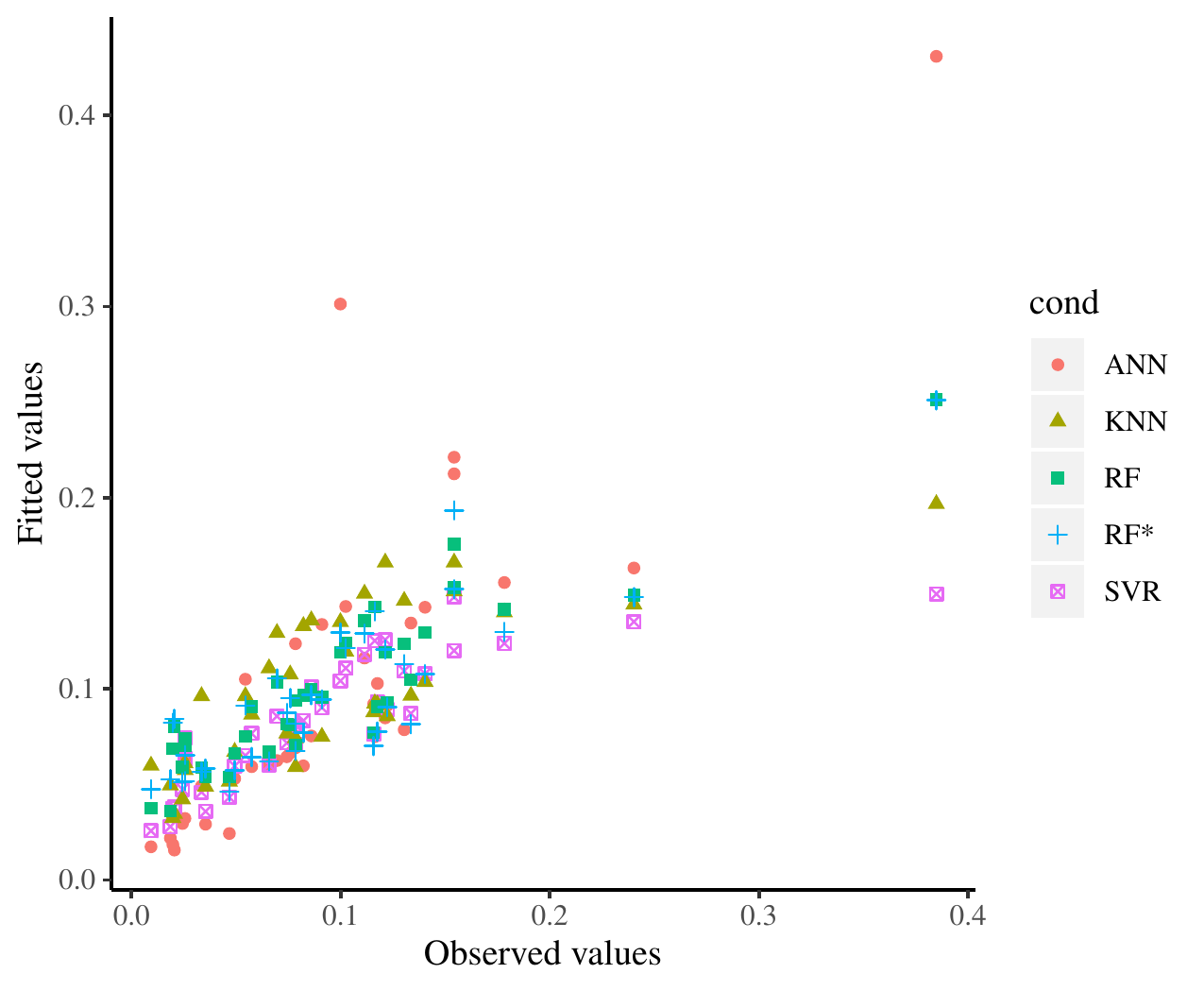}}
\subfigure[Batch 2]{\label{f:scatter2}  
\includegraphics[width=70mm]{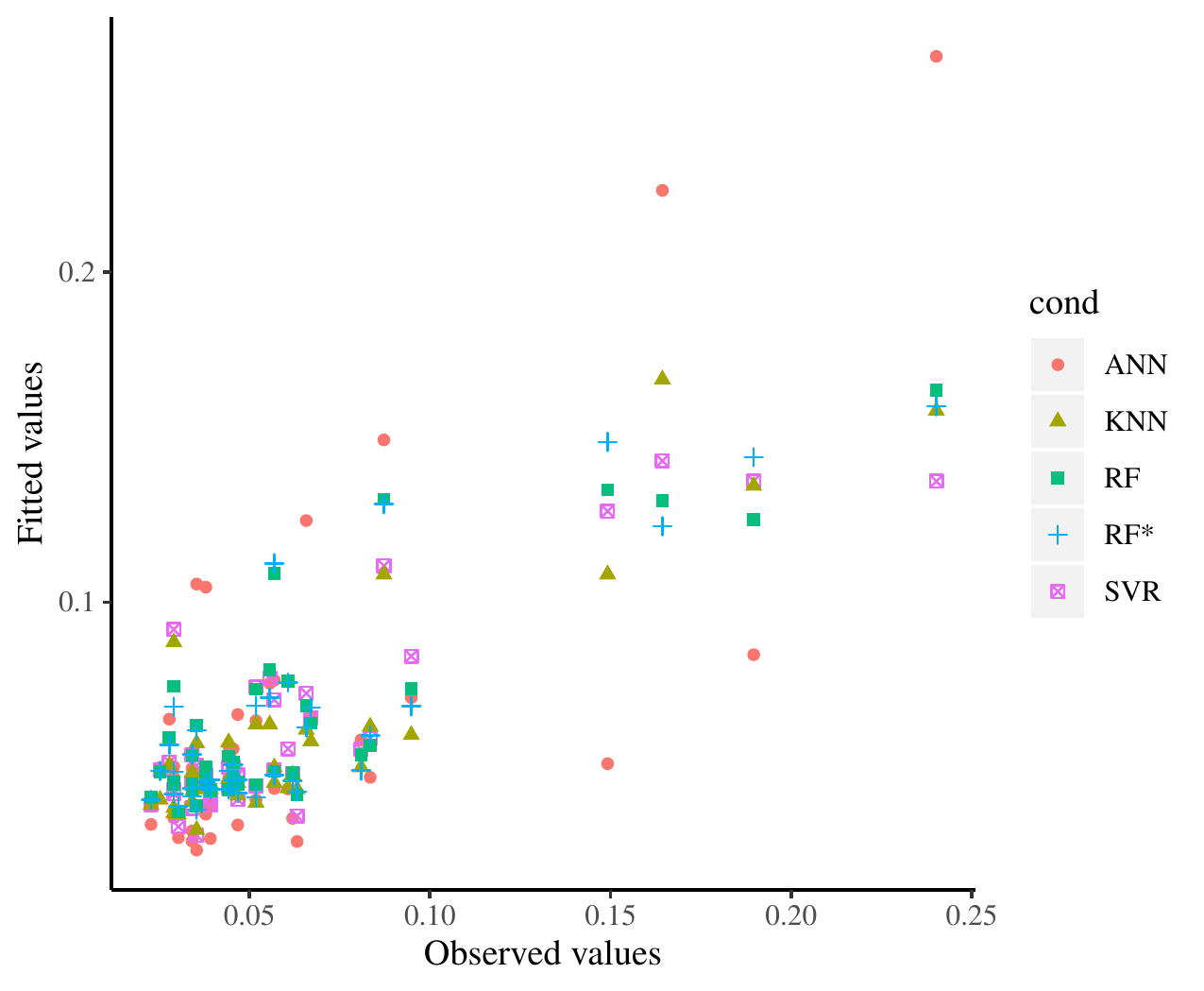}}
\subfigure[Batch 3]{\label{f:scatter3}  
\includegraphics[width=70mm]{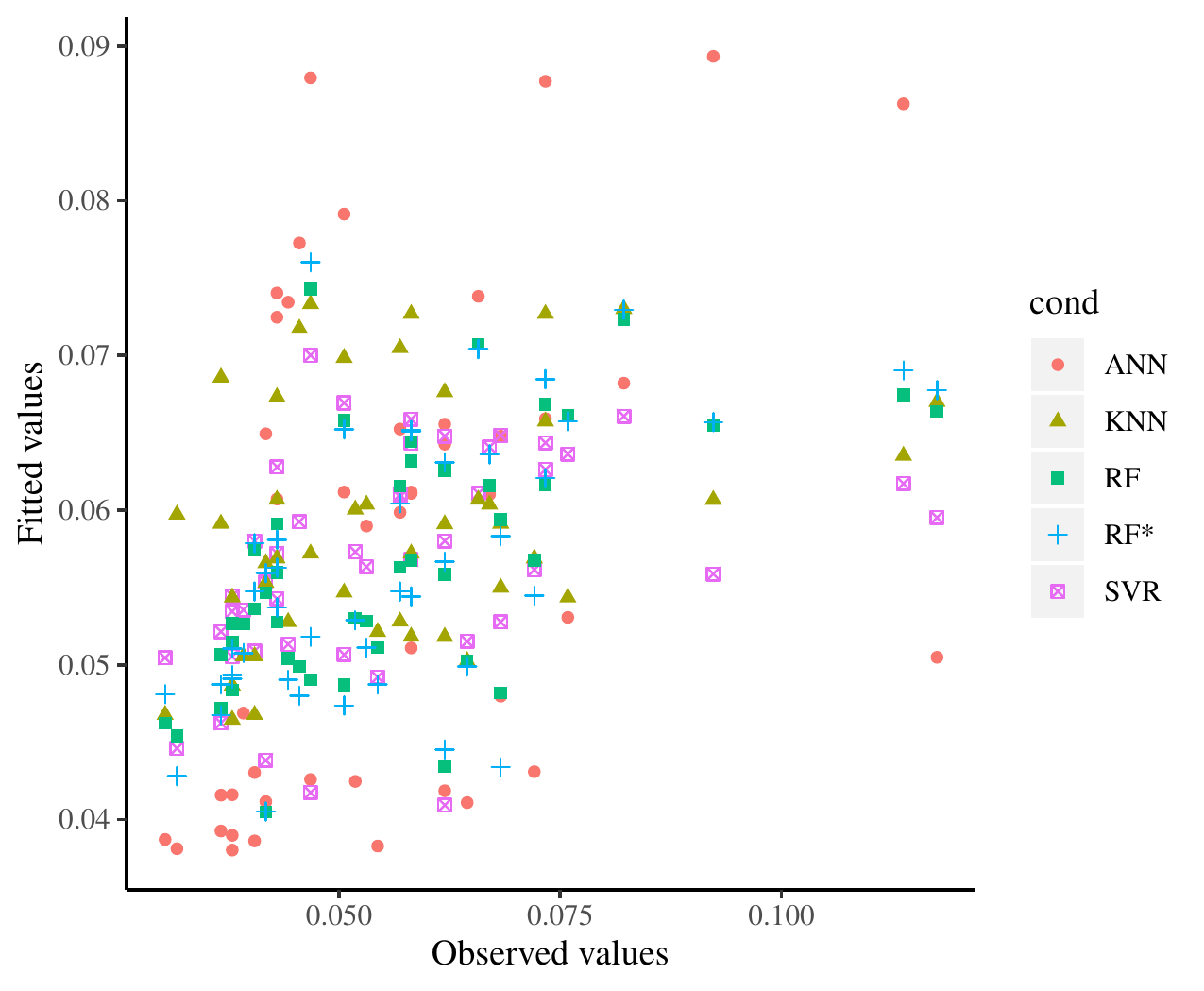}}
\caption{Scatterplot of observed and predicted values by means of different predictors: ANN, KNN, RF, RF* and SVR (see the text for more explanation).
}
\label{f:scater}
\end{figure}

Figure \ref{f:scater} shows a scatterplot of the actual and predicted female mosquito population values. In this plot, GLM, LM, and DTR are not considered due to their low RMSE figures. In all the Batches, the EO-based environmental features show significant effects on the vector population, less so in Batch 3. The environmental effects in Batches 2 and 3 show pointers with which the vector control program can be further improved. A plausible approach to consider is weighting the intensity of the program implementation by the most relevant environmental conditions across the year. This can ensure a higher control intensity during periods of favorable climate for vector development, thus reducing risks exposure during these periods and helping to optimize vector control resources allocation. As opposed to current common practice which favors the application of control actions mostly in rainy season, these actions should also be performed in dry season and periods of low vector population. Implementing control actions during the dry season has been shown in \cite{barsante2014model} to suppress the population of eggs laid during this season and consequently reduce the hatched population in the following rainy season.

%\begin{figure}
%\centering
%\subfigure[Year 2017]{\label{f:scatterv}  
%\includegraphics[width=0.45\textwidth]{scatterv-b}}
%\subfigure[Year 2018]{\label{f:scattert}  
%\includegraphics[width=0.45\textwidth]{scatterv-b-2018}}
%\caption{Scatterplot of observed and predicted values.
%}
%\label{f:scater}
%\end{figure}

In Table \ref{t:summaryt}, we present a summary of the observed and fitted data: mean, median, minimum (Min), maximum (Max), first (Q1) and third (Q3) quartiles. In Batch 1, LM exaggerates the minimum value,  producing a negative minimum value which has no physical meaning in relation to mosquito population. Considering that this same model produced a good quality measure of R, we deduce that LM, regardless of the quality score, is not a good model for predicting mosquito population because its distribution assumption of the response variable is inappropriate for this study use case. This problem, however, does not occur with GLM, due to its gamma distribution assumption which assumes positive values for $y$.  GLM produces a good estimated minimum value in all Batches, but still exhibits low prediction performance, which can be seen in its predicted mean and Q3 which are above the observed values for all Batches.  Among the ML methods, ANN produces better prediction in relation to minimum and maximum values.  RF and $\text{RF}^*$, though both overestimate the minimum in all Batches produced good estimated mean values and showed good balance.  The results in Figure \ref{f:scater}, Table \ref{t:quality-measures} and Table \ref{t:summaryt}  show that the best models for predicting the  vector population are $\text{RF}^*$ and RF, followed by SVR.

%\begin{table}%[ht]
%\centering
%\caption{Summary of the observed and fitted data}
%\label{t:summaryt}
%\begin{tabular}{rrrrrrr}
%  \hline
% & Min & Q1 & Median & Mean & Q3 & Max \\ 
%  \hline
% \multicolumn{7}{c}{Year 2017} \\
%\hline
%$y$ & 0.0247 & 0.1084 & 0.1590 & 0.1959 & 0.2473 & 0.8094 \\ 
%  GLM & 0.0002 & 0.0736 & 0.1362 & 0.2881 & 0.2211 & 4.1514 \\ 
%  LM & -1.4330 & 0.0164 & 0.1275 & 0.0798 & 0.2383 & 1.0131 \\ 
%  ANN & 0.0134 & 0.0587 & 0.1614 & 0.1954 & 0.2948 & 0.7091 \\ 
%  SVR & 0.0422 & 0.1225 & 0.1776 & 0.1753 & 0.2288 & 0.3080 \\ 
%  KNN & 0.0573 & 0.1383 & 0.2125 & 0.2058 & 0.2685 & 0.3562 \\ 
%  DTR & 0.0701 & 0.1188 & 0.2341 & 0.2119 & 0.2341 & 0.3499 \\ 
%  RF & 0.0949 & 0.1393 & 0.2015 & 0.2052 & 0.2495 & 0.4667 \\ 
%  $\text{RF}^*$ & 0.0877 & 0.1263 & 0.1835 & 0.1969 & 0.2342 & 0.5125 \\ 
%\hline
% \multicolumn{7}{c}{Year 2018} \\
%\hline
%$y$ & 0.0383 & 0.0747 & 0.0990 & 0.1151 & 0.1199 & 0.3344 \\ 
%  GLM & 0.0273 & 0.0626 & 0.0872 & 0.1135 & 0.1555 & 0.3527 \\ 
%  LM & -0.0800 & 0.0477 & 0.0966 & 0.1023 & 0.1515 & 0.3462 \\ 
%  ANN & 0.0431 & 0.0630 & 0.1014 & 0.1305 & 0.1529 & 0.4247 \\ 
%  SVR & 0.0638 & 0.0832 & 0.1030 & 0.1035 & 0.1163 & 0.1620 \\ 
%  KNN & 0.0538 & 0.0829 & 0.1067 & 0.1085 & 0.1305 & 0.1875 \\ 
%  DTR & 0.0864 & 0.0864 & 0.1086 & 0.1185 & 0.1592 & 0.1686 \\ 
%  RF & 0.0778 & 0.0920 & 0.1135 & 0.1182 & 0.1299 & 0.2740 \\ 
%  $\text{RF}^*$ & 0.0786 & 0.0918 & 0.1085 & 0.1173 & 0.1258 & 0.2849 \\ 
%\hline
%\end{tabular}
%\end{table}

\begin{table}%[ht]
\footnotesize
\centering
\caption{Summary of the observed and fitted data}
\label{t:summaryt}
\begin{tabular}{rrrrrrr}
  \hline
 & Min & Q1 & Median & Mean & Q3 & Max \\ 
  \hline
 \multicolumn{7}{c}{Batch 1} \\
\hline
 $y$ & 0.0093 & 0.0474 & 0.0803 & 0.0933 & 0.1204 & 0.3848 \\ 
  GLM & 0.0063 & 0.0397 & 0.0745 & 0.1043 & 0.1611 & 0.4340 \\ 
  LM & -0.0724 & 0.0413 & 0.0793 & 0.0946 & 0.1330 & 0.3565 \\ 
  ANN & 0.0157 & 0.0546 & 0.0752 & 0.0995 & 0.1312 & 0.4309 \\ 
  SVR & 0.0257 & 0.0610 & 0.0822 & 0.0838 & 0.1090 & 0.1496 \\ 
  KNN & 0.0325 & 0.0621 & 0.0928 & 0.0984 & 0.1357 & 0.1967 \\ 
  DTR & 0.0392 & 0.0392 & 0.0741 & 0.0986 & 0.1411 & 0.1715 \\ 
  RF & 0.0362 & 0.0703 & 0.0918 & 0.0985 & 0.1224 & 0.2512 \\ 
  $\text{RF}^*$ & 0.0461 & 0.0646 & 0.0858 & 0.0956 & 0.1187 & 0.2510 \\ 
 \hline
 \multicolumn{7}{c}{Batch 2} \\
\hline
 $y$ & 0.0228 & 0.0351 & 0.0461 & 0.0619 & 0.0638 & 0.2402 \\ 
  GLM & 0.0208 & 0.0384 & 0.0695 & 0.0890 & 0.0976 & 0.5542 \\ 
  LM & 0.0140 & 0.0376 & 0.0716 & 0.0795 & 0.0977 & 0.3030 \\ 
  ANN & 0.0247 & 0.0370 & 0.0488 & 0.0694 & 0.0857 & 0.2653 \\ 
  SVR & 0.0309 & 0.0456 & 0.0498 & 0.0614 & 0.0703 & 0.1351 \\ 
  KNN & 0.0310 & 0.0450 & 0.0487 & 0.0595 & 0.0566 & 0.1688 \\ 
  DTR & 0.0438 & 0.0438 & 0.0438 & 0.0618 & 0.0641 & 0.1439 \\ 
  RF & 0.0370 & 0.0454 & 0.0500 & 0.0642 & 0.0701 & 0.1769 \\ 
  $\text{RF}^*$ & 0.0367 & 0.0436 & 0.0507 & 0.0639 & 0.0678 & 0.1856 \\ 
  \hline
 \multicolumn{7}{c}{Batch 3} \\
\hline
  $y$ & 0.0303 & 0.0417 & 0.0531 & 0.0563 & 0.0657 & 0.1176 \\ 
  GLM & 0.0303 & 0.0440 & 0.0576 & 0.0606 & 0.0702 & 0.1020 \\ 
  LM & 0.0251 & 0.0453 & 0.0598 & 0.0596 & 0.0693 & 0.0888 \\ 
  ANN & 0.0380 & 0.0419 & 0.0599 & 0.0574 & 0.0659 & 0.0893 \\ 
  SVR & 0.0409 & 0.0515 & 0.0568 & 0.0566 & 0.0626 & 0.0700 \\ 
  KNN & 0.0389 & 0.0544 & 0.0591 & 0.0592 & 0.0657 & 0.0777 \\ 
  DTR & 0.0478 & 0.0478 & 0.0584 & 0.0576 & 0.0673 & 0.0673 \\ 
  RF & 0.0405 & 0.0504 & 0.0558 & 0.0565 & 0.0626 & 0.0743 \\ 
  $\text{RF}^*$ & 0.0405 & 0.0491 & 0.0547 & 0.0562 & 0.0636 & 0.0760 \\ 
\hline
\end{tabular}
\end{table}

\begin{figure}
\centering
\subfigure[Batch 1]{\label{f:predictionv}  
\includegraphics[width=0.38\textwidth]{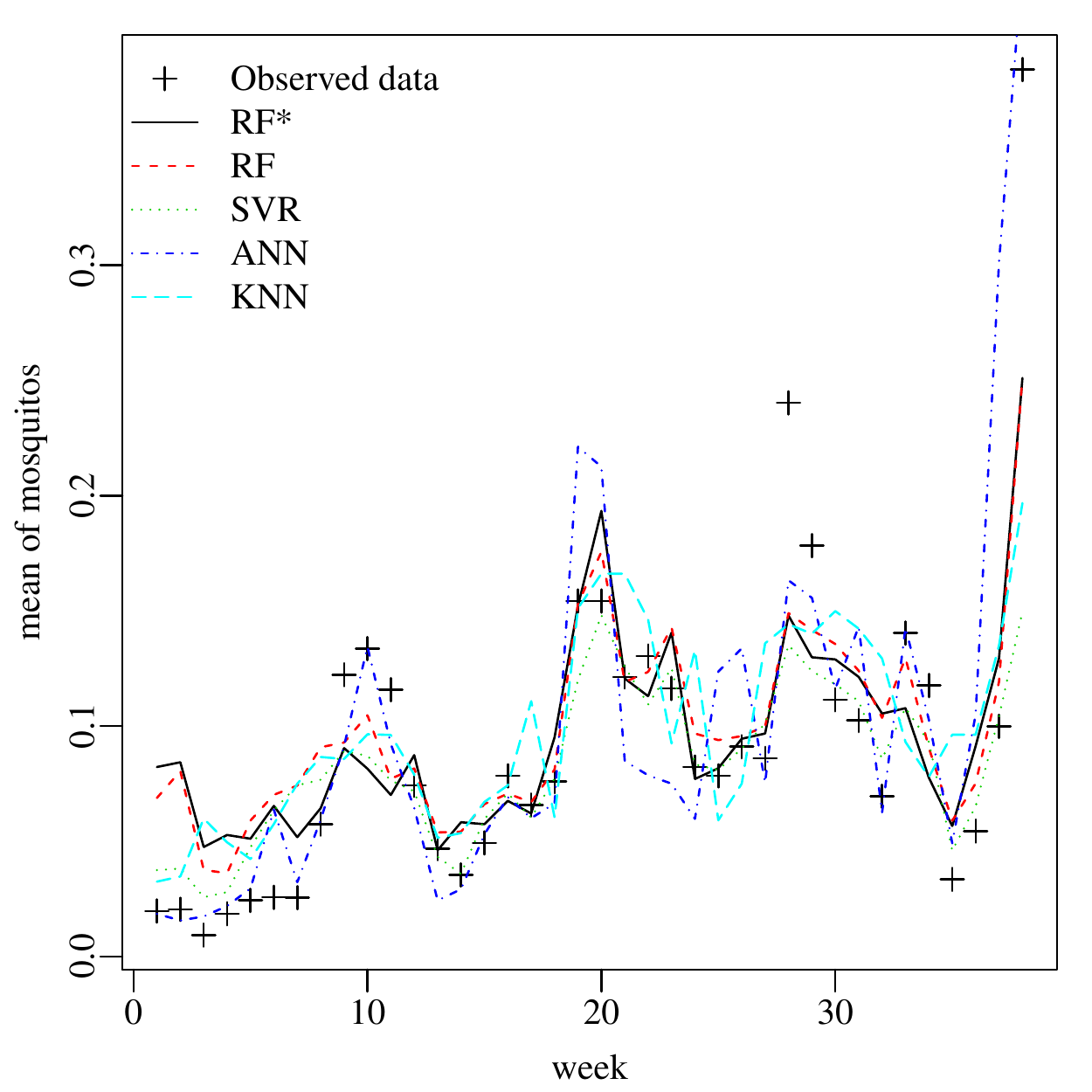}}
\subfigure[Batch 2]{\label{f:predictiont}  
\includegraphics[width=0.38\textwidth]{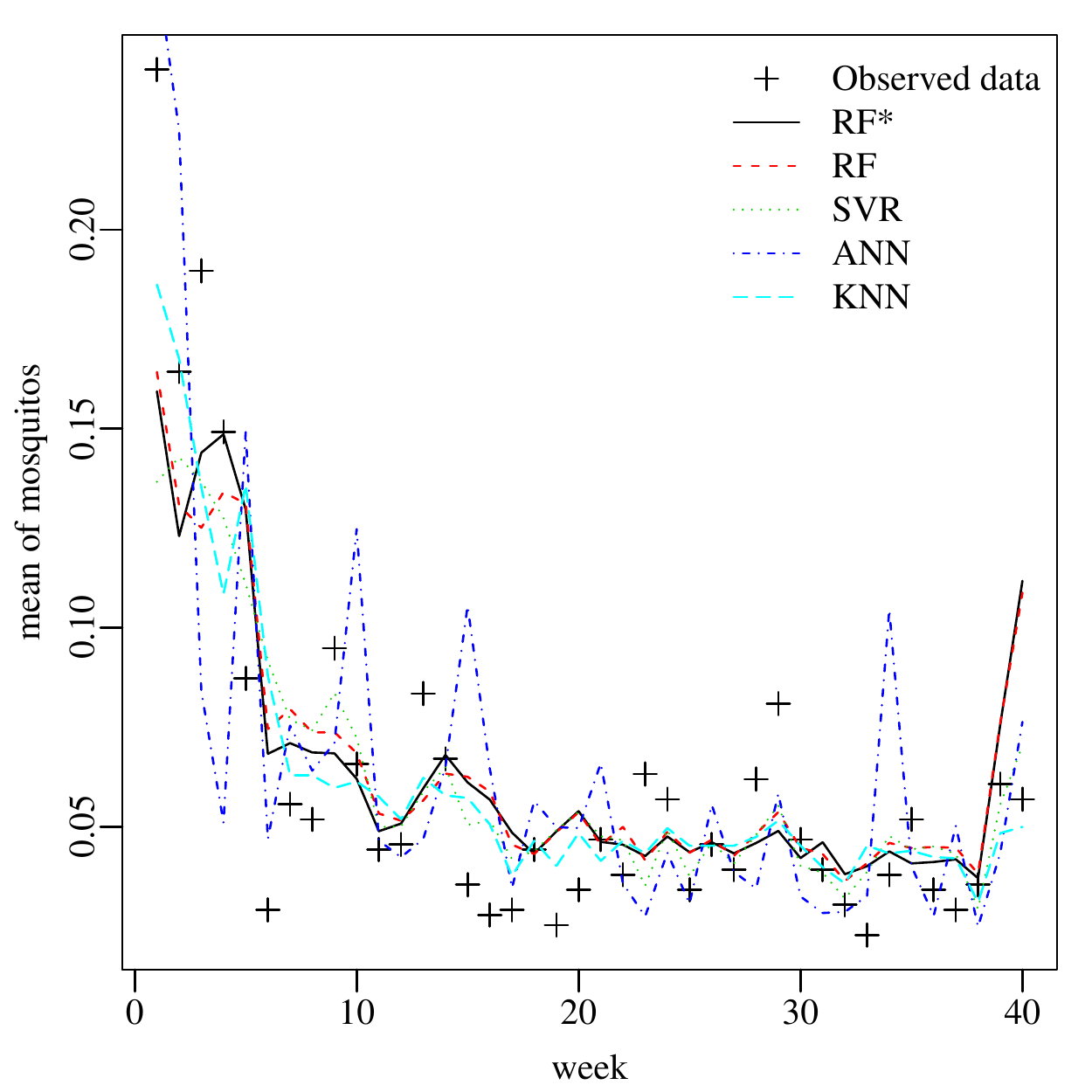}}
\subfigure[Batch 3]{\label{f:predictiont}  
\includegraphics[width=0.38\textwidth]{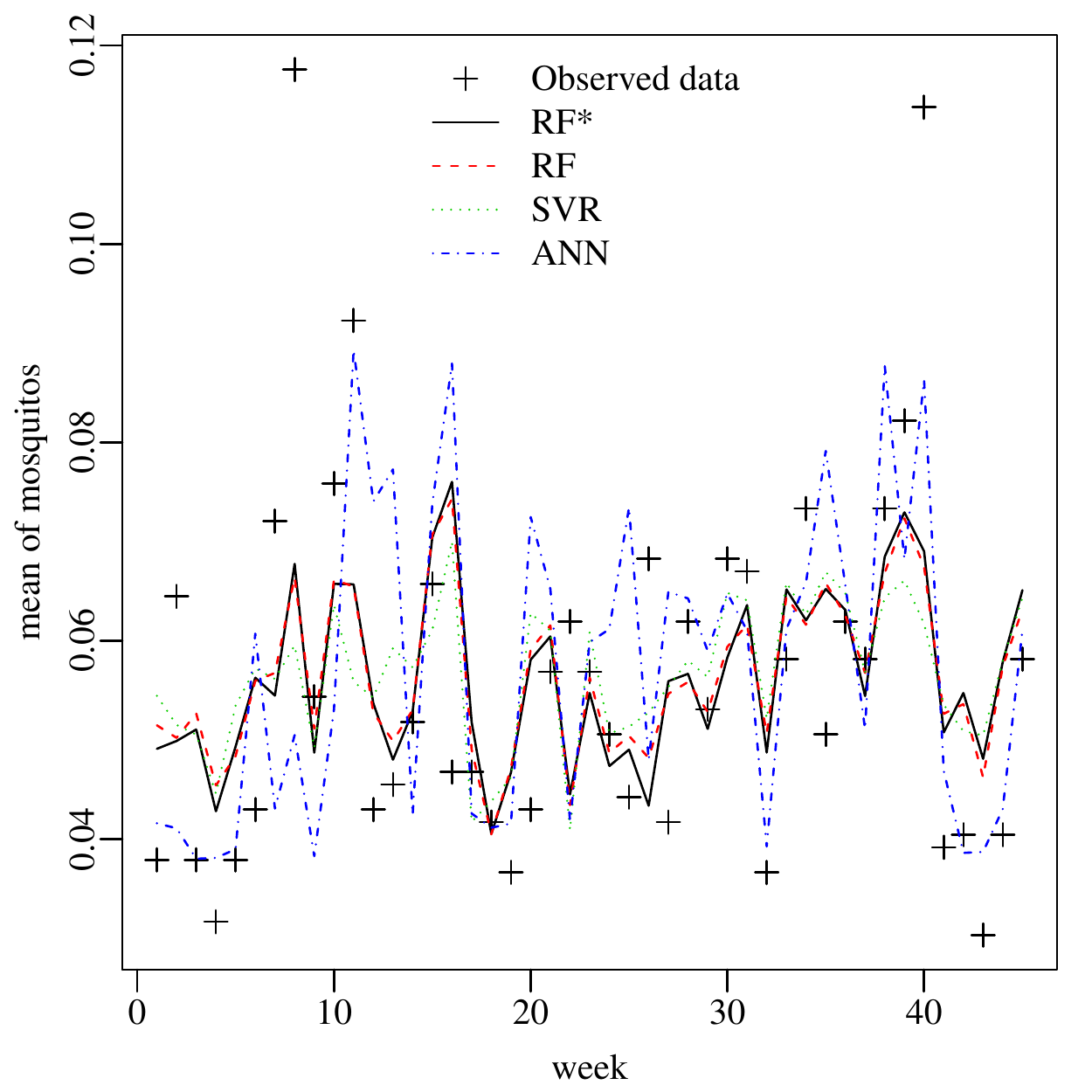}}
\caption{ Time series of the actual  (points) versus predicted (lines) values,using only a subset of the predition algorithms in Fig. 4: ANN, SVR, RF and $\text{RF*}$. 
}
\label{f:prediction}
\end{figure}

Figure \ref{f:prediction} shows the mosquito population data (validation set) along with the predicted values by the best models. It is seen that the SVR and KNN models underestimate  the  seasonal spikes in Batches 1 and 2.  The model performances are reduced in Batches 2 and 3 to their overestimation of the very low mosquito population value points resulting from the improved control activities in the Batch periods.

\begin{figure}
\centering
\subfigure[Batch 1]{\label{f:mdi17}  
\includegraphics[width=0.38\textwidth]{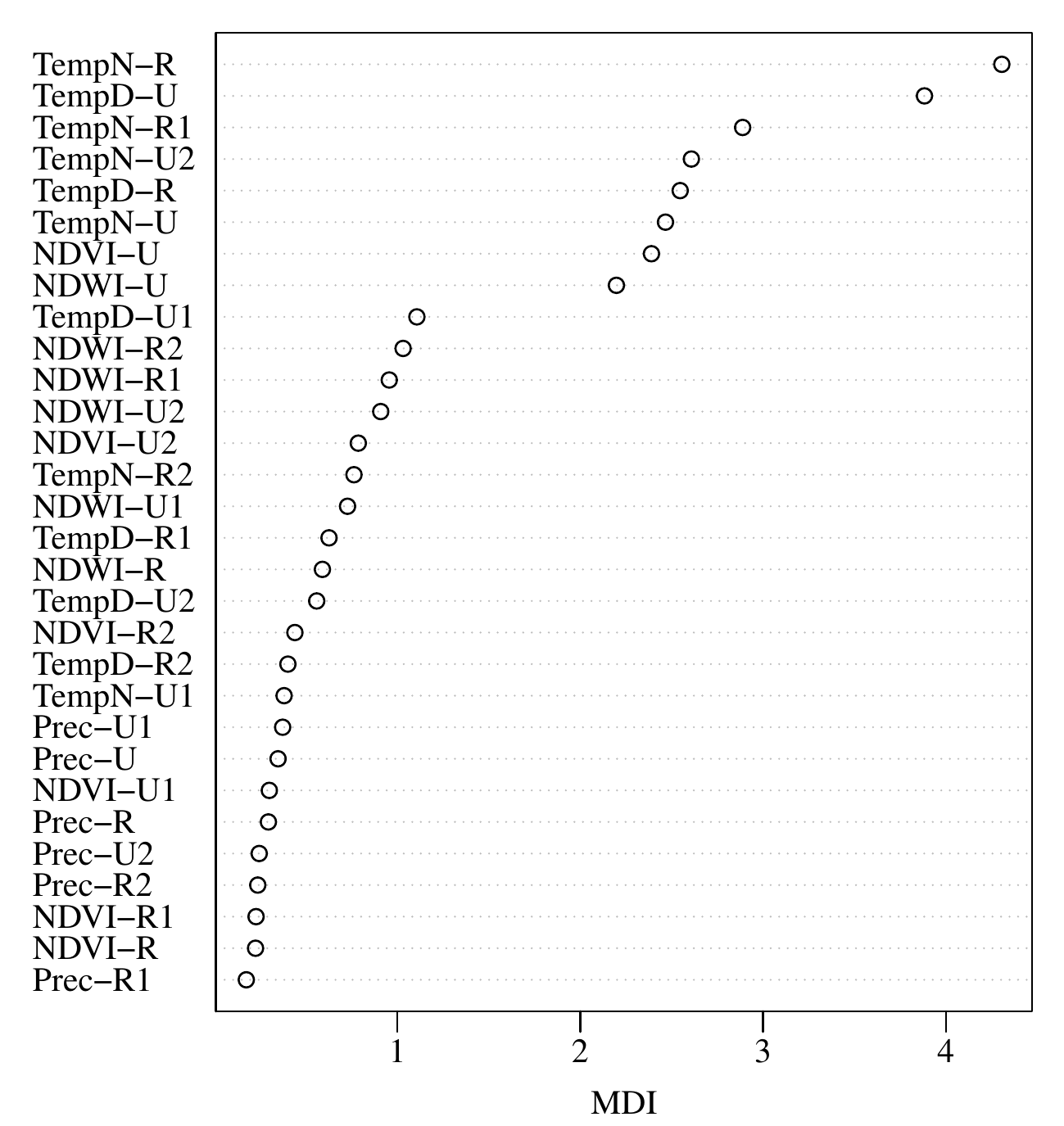}}
\subfigure[Batch 2]{\label{f:mdi18}  
\includegraphics[width=0.38\textwidth]{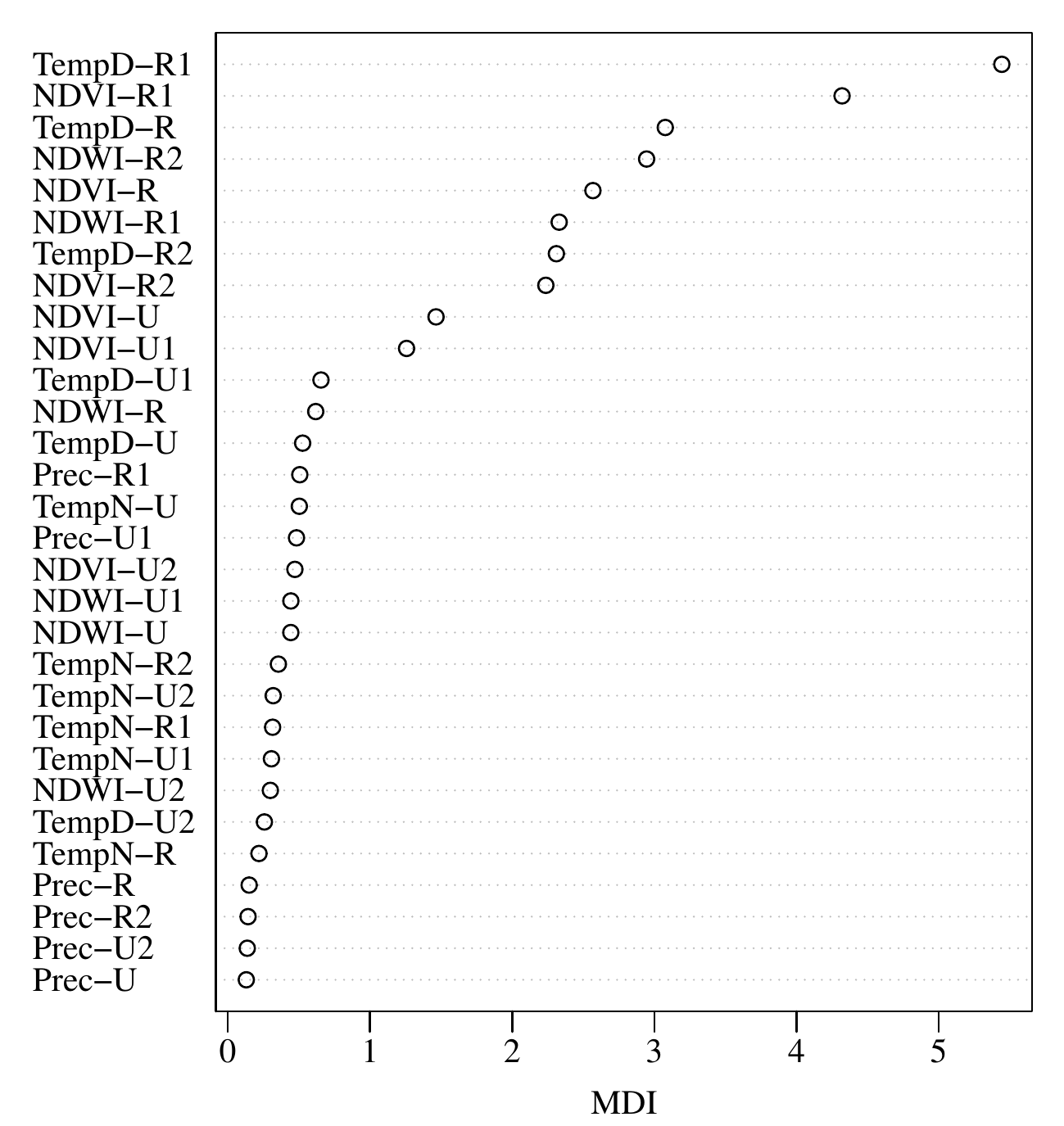}}
\subfigure[Batch 3]{\label{f:mdi19}  
\includegraphics[width=0.38\textwidth]{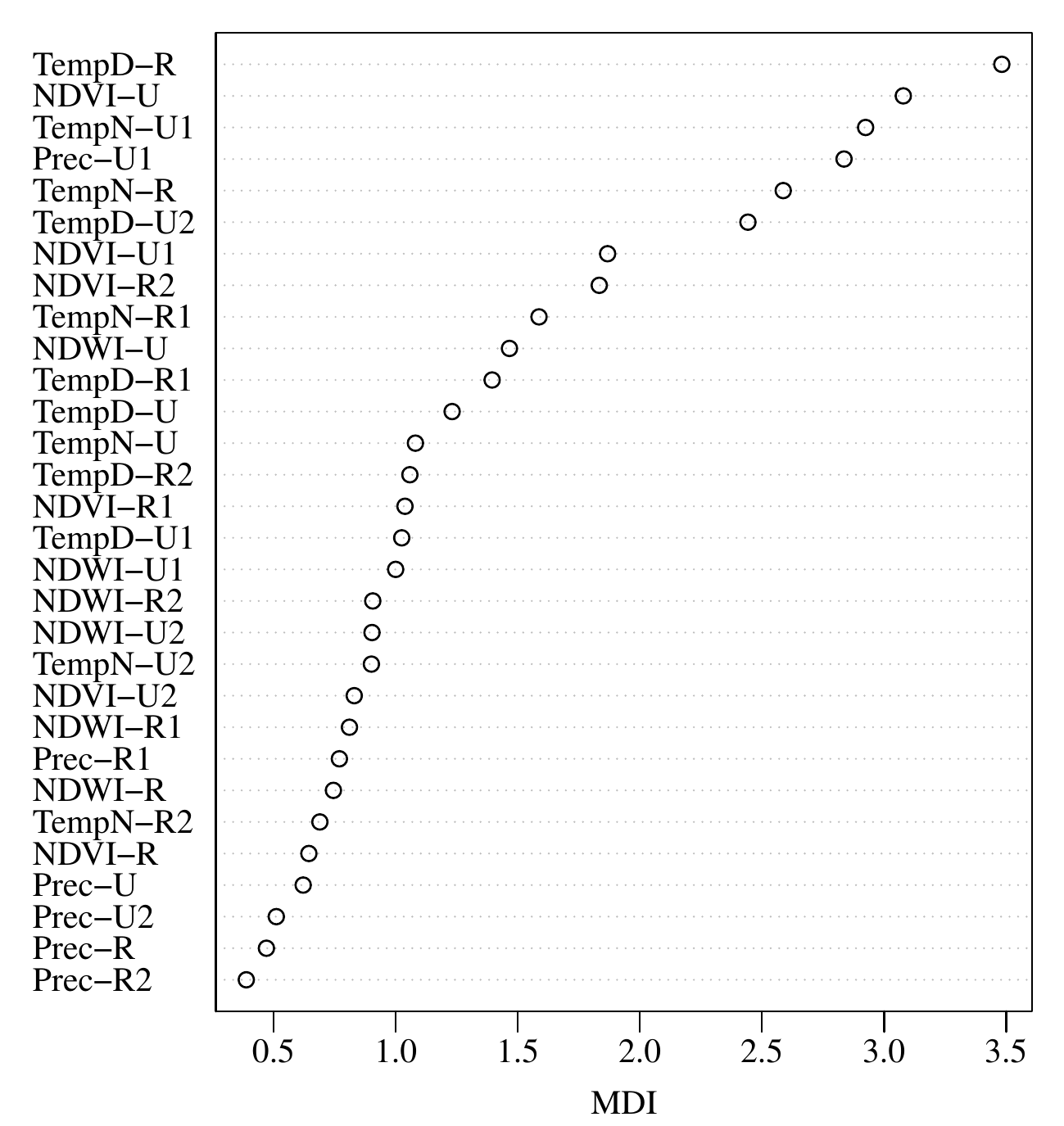}}
\caption{Average values of MDI considering 50 RF replicas.}
\label{f:importance-rf}
\end{figure}

In further analysis of $\text{RF*}$, here, we discuss the selected most informative features subset across the considered observation time periods. Figure \ref{f:importance-rf} shows the average value of the MDI for each variable considering 50 RF replicas. For Batch 1, the eight selected most informative EO variables are TempN-R, TempD-U, TempN-R1, TempN-U2, TempD-R, TempN-U, NDVI-U and NDWI-U. For Batch 2, the selected variables features are: TempD-R1, NDVI-R1, TempD-R, NDWI-R2, NDVI-R, NDWI-R1, TempD-R2 and NDWI-R2. For Batch 3, they are: TempD-R, NDVI-U, TempN-U1, Prec-U1, TempN-R, TempD-U2, NDVI-U1, NDVI-R2. The improved control actions in Batches 2 and 3 are mostly responsible for the differences in important variables across all Batches. Only TempD-R is commonly selected in all the Batches. In addition, we can see that  temperature features extracted from the rural surface zone provide the highest quality of information to our models in all Batches: TempN-R, TempD-R1 and TempD-R for Batches 1, 2 and 3, respectively. This is consistent with previous studies which show that non-artificial surface characterizations of the environment are more informative for predicting \textit{Ae. aegypti} vector population \cite{Scavuzzo2018, German2018}, and that the weekly (or daily) temperature is the most important environmental condition affecting the development of \textit{Ae. aegypti} \cite{cheong2013assessing}. The non-synchronous effects of temperature is also seen in the selection of both TempN-R and  TempN-R1 in Batch 1, and TempD-R, TempD-R1 and TempD-R2 in Batch 2. Of all eight selected subset variables, six of them lagged in both Batches 2 and 3, compared to only two in Batch 1. This shows that most of the environmental effects on vector population during the improved vector control regimes come from non-synchronous compounding effects. Six temperature variables are selected in Batch 1. This number reduced to three in both Batches 2 and 3. We deduce from this that the effect of non-temperature environmental variables increases during the improved control actions regime.

\begin{figure}
\centering
\subfigure[Batch 1]{\label{f:relation17}  
\includegraphics[width=0.38\textwidth]{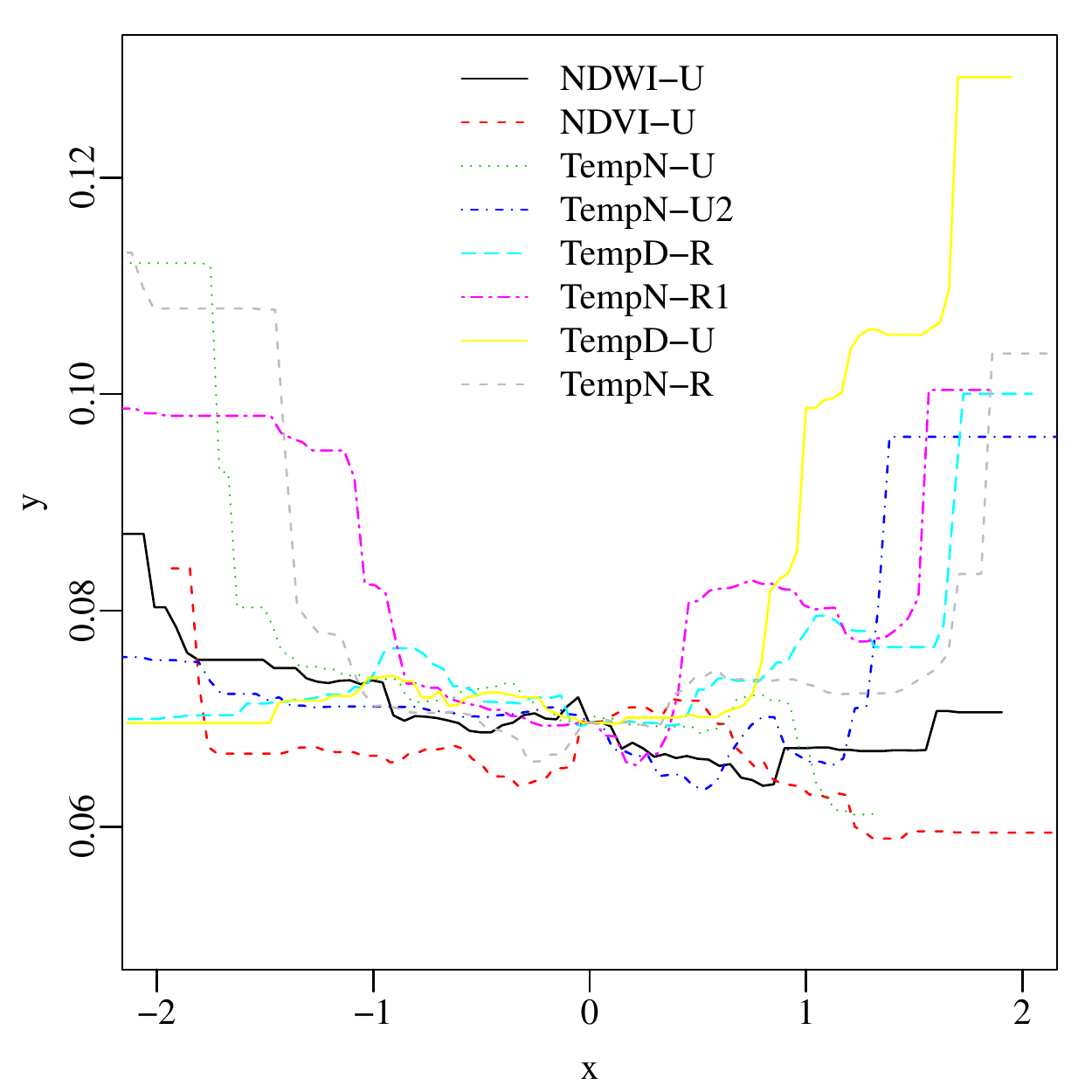}}
\subfigure[Batch 2]{\label{f:relation18}  
\includegraphics[width=0.38\textwidth]{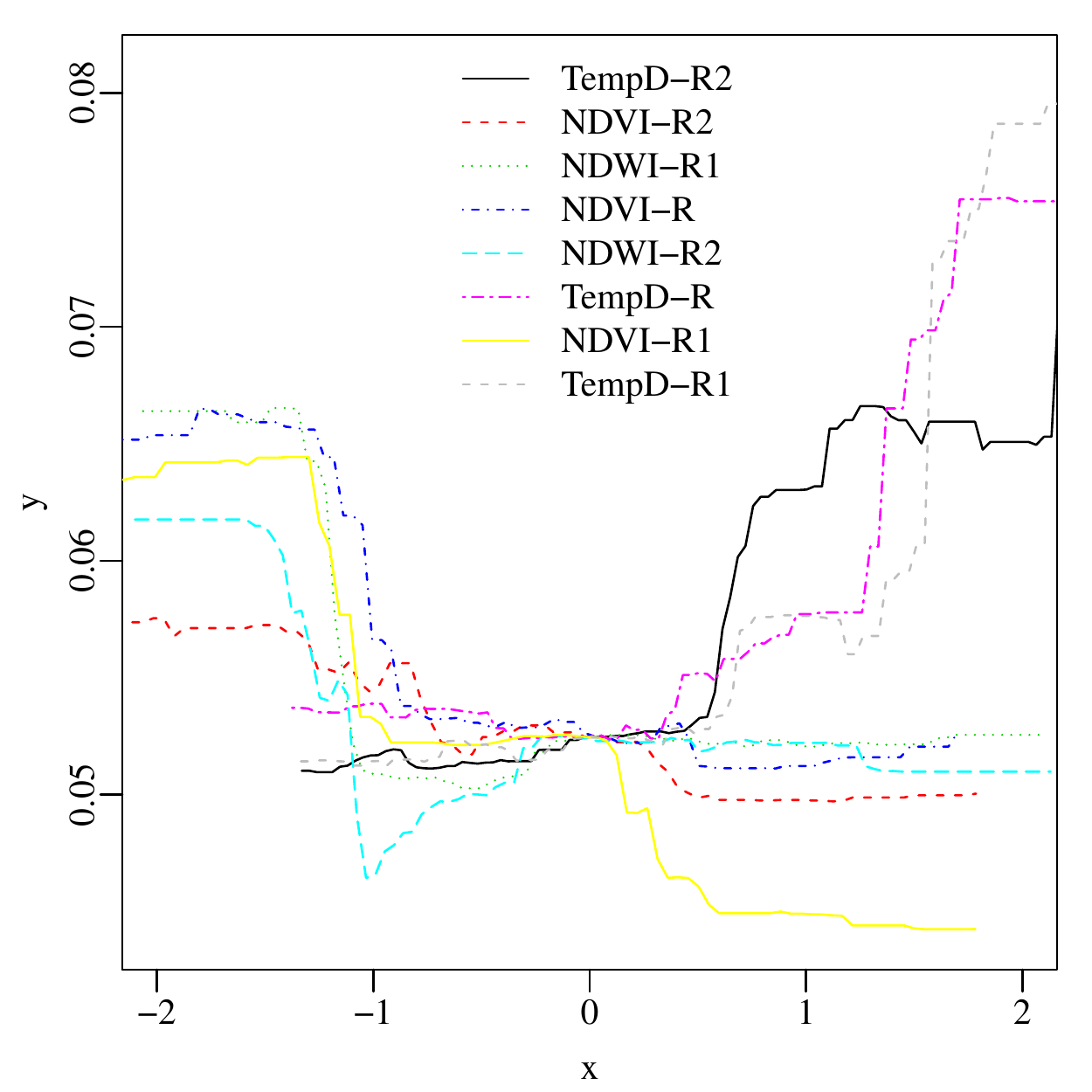}}
\subfigure[Batch 3]{\label{f:relation19}  
\includegraphics[width=0.38\textwidth]{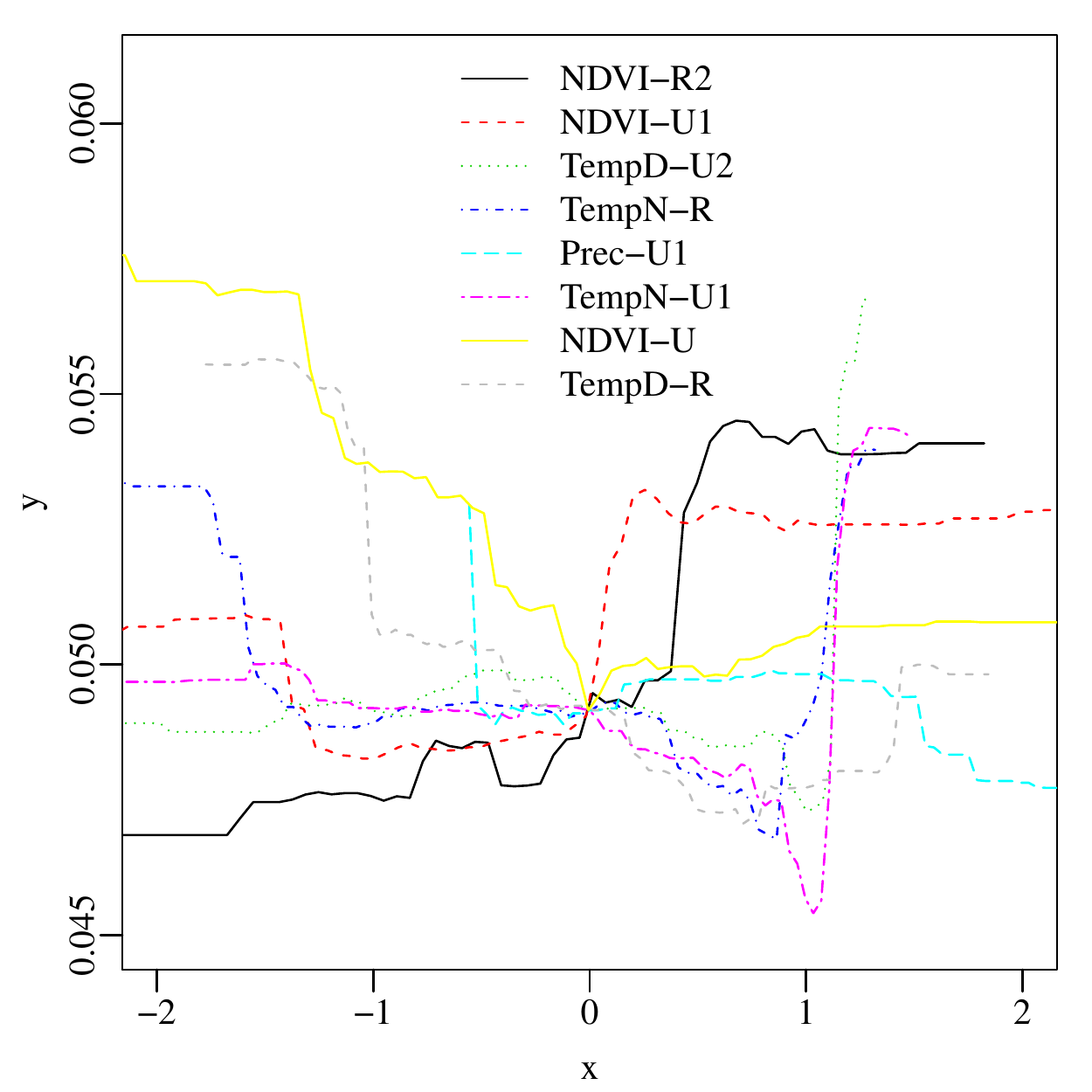}}
\caption{The relation between the mean of mosquitoes ($y$) and each covariate ($x$).}
\label{f:relation}
\end{figure}

Figure \ref{f:relation} presents the relationship between the mosquito population $y$ and the standardized values of environmental features selected for $\text{RF}^*$.   It is seen that the relationships vary across the different features ($x$). For example,  Figure \ref{f:relation17}  shows that in Batch 1, higher values of all six selected temperature variables,  between their mean  and two standard deviations above their mean, result in larger numbers of mosquitoes. Also, for Batches 2 and 3, as shown in Figures \ref{f:relation18} and \ref{f:relation19}, even with the improved anthropological interference due to better control actions, when the three selected most informative temperature variables in each case synchronously rise above one standard deviation from their mean, there is a rapid increase in the vector population.   These results are in accordance with laboratory studies presented in \cite{Jansen2010, kamimura2002effect}.  These works show that at higher temperatures below 40 \textcelsius,  the development life cycle of \textit{Ae. aegypti} is accelerated,  thus increasing the vector population. In addition, the extrinsic incubation period (EIP) of \textit{Ae. aegypti} -- the time interval between virus agent acquisition by the mosquito vector and the moment it is able to transmit it to humans -- reduces at high temperature \cite{Liu2017, Tjaden2013}. The results in Figure \ref{f:relation18} also show that the improved control program in starting from  Batch 2 is effective except in cases of synchronous extreme values of important temperature variables. This observation shows pointers that can be used to further improve the efficacy of the control actions in subsequent years.

Regarding non-temperature environmental effects, the relation curves for Batch 3, Figure \ref{f:relation19}, also show that  at about $-$0.5 standard deviation below the mean of Prec-U1, there is a spike in the vector population. As reported in \cite{Jansen2010, Scavuzzo2018}, in regions where primary larval sites are in rain-filled containers, rainfall has been shown to positively correlate with larval and adult mosquito abundance.  Further studies have  corroborated the importance of humidity and vegetation conditions in abundance and reproduction of mosquito species, with longer dry season and lower relative humidity resulting in higher egg mortality \cite{trpivs1972dry}.  Fully developed tree canopies by providing shade can reduce evaporation of hatching water, and can also increase near ground humidity, thus increasing density of \textit{Ae. aegypti} mosquito larvae \cite{Messina2016, fuller2009nino}. Stronger positive effects of vegetation conditions can be observed through the effects of NDVI-R2 and NDVI-U1 in Batch 3.

Finally, and in general, this research shows, based on our study data, that the random forest model performs better than all the other models for predicting the mosquito population from EO-based variables.  By selecting the most informative features using the MDI measure, it is possible to obtain a much less complex model with comparable (or even better) results, labeled in this work as $\text{RF*}$. 
Additionally, with the MDI ranking it is possible to understand and explain the rationale for the model variables and analyse the effects of each EO variable on the vector population. This better understanding is gained by considering the relation graphs based on $\text{RF*}$, which provide a visualization of the the nonlinear relationships among the variables. 

\section{Conclusions} 
In this paper, a procedure for modeling the temporal population distribution of female \textit{Ae. aegypti} mosquito species based effects environmental factors estimated from freely available EO data has been presented. The procedure was implemented using MODIS and GPM EO data products to obtain information about temperature, precipitation, moisture and vegetation conditions. The mosquito vector population data was collected on the field from April 2017 to August 2019 (124 weeks) using 791 mosquito traps. We divided the field collected data in time into three operational Batches based on the difference in data collection conditions as obtained from the field staff. From January 2018, there were data-driven improvements to vector control program, affecting data Batches 2 and 3 in our study. 

Different studies in the recent past have been conducted in this domain due to the rise in availability of more freely available earth observation data and off-the-shelf machine learning analyses tools \cite{Scavuzzo2018, German2018}. These studies, being prediction focused, do not consider explainable modeling using non-linear machine learning techniques like RF. 

For this reason, here, we have presented a procedure for explainable modeling of female \textit{Ae. aegypti} based on EO data estimated environmental factors based on RF. A quantitative measure of the variable importance (MDI)--- wrapped in RF --- was used to extract the most informative environmental features, and obtain a less complex but still accurate predictive RF model, labeled $\text{RF*}$ in this study. To prove their robustness, the resulting models were compared to other machine learning models including SVR, ANN, KNN and DTR, as well as statistical models, such as LM and GLM.

Our results show that our proposed RF-based approach is capable of mapping the complex relationship among the EO variables and vector population. Furthermore, the features selected thanks to the MDI value ranking can be empirically interpreted and provide hints about the relationships among vector population and these environmental conditions from an operational point of view.

\section*{Acknowledgment} 
The first author gratefully acknowledges the financial support by the European Commission through Horizon 2020 research and innovation programme, grant agreement No 734541 - Project "EOXPOSURE".  FMB thanks to CNPq/MCTI, Brazil.  AEE thanks CAPES/MEC, CNPq/MCTI and Decit/SCTIE/MS (grant No 440358/2016-7). All the authors also thank Ecovec for providing the mosquito trap data and Marcelaine Marculano (Vigilancia Ambiental) for the entomological data.  

\bibliographystyle{unsrt}  

  %%% Remove comment to use the external .bib file (using bibtex).


\begin{thebibliography}{10}

\bibitem{Linard2013}
Catherine Linard, Andrew~J. Tatem, and Marius Gilbert.
\newblock Modelling spatial patterns of urban growth in {A}frica.
\newblock {\em Applied Geography}, 44:23--32, 2013.

\bibitem{donnay2000remote}
Jean-Paul Donnay, Mike~J Barnsley, and Paul~A Longley.
\newblock {\em Remote Sensing and Urban Analysis: GISDATA 9}.
\newblock CRC Press, 2000.

\bibitem{Bagan2012}
Hasi Bagan and Yoshiki Yamagata.
\newblock Landsat analysis of urban growth: How {T}okyo became the
  world{\textquotesingle}s largest megacity during the last 40years.
\newblock {\em Remote Sensing of Environment}, 127:210--222, 2012.

\bibitem{QihaoGamba2018a}
Paolo E.~Gamba Qihao~Weng, Dale~Quattrochi.
\newblock {\em Urban Remote Sensing, Second Edition}.
\newblock {CRC} Press, 2018.

\bibitem{Weng2014}
Qihao Weng.
\newblock {\em Global Urban Monitoring and Assessment through Earth
  Observation}.
\newblock CRC Press, 2014.

\bibitem{While2013}
Aidan While and Mark Whitehead.
\newblock Cities, urbanisation and climate change.
\newblock {\em Urban Studies}, 50(7):1325--1331, 2013.

\bibitem{Markham2012}
Brian~L. Markham and Dennis~L. Helder.
\newblock Forty-year calibrated record of {E}arth-reflected radiance from
  {L}andsat: A review.
\newblock {\em Remote Sensing of Environment}, 122:30--40, 2012.

\bibitem{patel2015multitemporal}
Nirav~N Patel, Emanuele Angiuli, Paolo Gamba, Andrea Gaughan, Gianni Lisini,
  Forrest~R Stevens, Andrew~J Tatem, and Giovanna Trianni.
\newblock Multitemporal settlement and population mapping from {L}andsat using
  {G}oogle {E}arth {E}ngine.
\newblock {\em International Journal of Applied Earth Observation and
  Geoinformation}, 35:199--208, 2015.

\bibitem{jiang2017spatial}
Zhe Jiang and Shashi Shekhar.
\newblock {\em Spatial Big Data Science}.
\newblock Springer, 2017.

\bibitem{Marinoni2017}
Andrea Marinoni and Paolo Gamba.
\newblock Unsupervised data driven feature extraction by means of mutual
  information maximization.
\newblock {\em {IEEE} Transactions on Computational Imaging}, 3(2):243--253,
  2017.

\bibitem{Turyahikayo2015}
Agnes Turyahikayo.
\newblock Assessing land use induced disturbance to vegetation cover in the
  upper {M}olopo catchment, {S}outh {A}frica, using {L}andsat images.
\newblock In {\em 2015 {IEEE} International Geoscience and Remote Sensing
  Symposium ({IGARSS})}. {IEEE}, 2015.

\bibitem{OSTFELD2005}
R.~Ostfeld, G.~Glass, and F~Keesing.
\newblock Spatial epidemiology: an emerging (or re-emerging) discipline.
\newblock {\em Trends in Ecology {\&} Evolution}, 20(6):328--336, 2005.

\bibitem{Reisen2010}
William~K. Reisen.
\newblock Landscape epidemiology of vector-borne diseases.
\newblock {\em Annual Review of Entomology}, 55(1):461--483, 2010.

\bibitem{Young2013}
Sean~G. Young, Jason~A. Tullis, and Jackson Cothren.
\newblock A remote sensing and {GIS}-assisted landscape epidemiology approach
  to {W}est {N}ile virus.
\newblock {\em Applied Geography}, 45:241--249, 2013.

\bibitem{Rossi2018}
Jean-Pierre Rossi, Ibrahima Kadaour{\'{e}}, Martin Godefroid, and Gauthier
  Dobigny.
\newblock Landscape epidemiology in urban environments: The example of
  rodent-borne trypanosoma in {N}iamey, {N}iger.
\newblock {\em Infection, Genetics and Evolution}, 63:307--315, 2018.

\bibitem{Johnson2017}
Tammi~L Johnson, Ubydul Haque, Andrew~J Monaghan, Lars Eisen, Micah~B Hahn,
  Mary~H Hayden, Harry~M Savage, Janet McAllister, John-Paul Mutebi, and
  Rebecca~J Eisen.
\newblock Modeling the environmental suitability for {A}edes (stegomyia)
  aegypti and {A}edes (stegomyia) albopictus (diptera: Culicidae) in the
  contiguous {U}nited {S}tates.
\newblock {\em Journal of Medical Entomology}, 54(6):1605--1614, 2017.

\bibitem{Fornace2014}
Kimberly~M. Fornace, Chris~J. Drakeley, Timothy William, Fe~Espino, and
  Jonathan Cox.
\newblock Mapping infectious disease landscapes: unmanned aerial vehicles and
  epidemiology.
\newblock {\em Trends in Parasitology}, 30(11):514--519, 2014.

\bibitem{Jamison2015}
Amanda Jamison, Elaina Tuttle, Ryan Jensen, Greg Bierly, and Rusty Gonser.
\newblock Spatial ecology, landscapes, and the geography of vector-borne
  disease: A multi-disciplinary review.
\newblock {\em Applied Geography}, 63:418--426, 2015.

\bibitem{rotela2017analytical}
Camilo Rotela, Laura Lopez, Mar{\'\i}a~Fr{\'\i}as C{\'e}spedes, Gabriela
  Barbas, Andr{\'e}s Lighezzolo, Ximena Porcasi, Mario~A Lanfri, Carlos~M
  Scavuzzo, and David~E Gorla.
\newblock Analytical report of the 2016 dengue outbreak in {C}{\'o}rdoba city,
  {A}rgentina.
\newblock {\em Geospatial {H}ealth}, 2017.

\bibitem{Porcasi2012}
Ximena Porcasi, Camilo~H. Rotela, Mar{\'{\i}}a~V. Introini, Nicol{\'{a}}s
  Frutos, Sof{\'{\i}}a Lanfri, Gonzalo Peralta, Estefan{\'{\i}}a A.~De Elia,
  Mario~A. Lanfri, and Carlos~M. Scavuzzo.
\newblock An operative dengue risk stratification system in {A}rgentina based
  on geospatial technology.
\newblock {\em Geospatial Health}, 6(3):31, 2012.

\bibitem{Messina2016}
Jane~P Messina, Moritz~UG Kraemer, Oliver~J Brady, David~M Pigott, Freya~M
  Shearer, Daniel~J Weiss, Nick Golding, Corrine~W Ruktanonchai, Peter~W
  Gething, Emily Cohn, John~S Brownstein, Kamran Khan, Andrew~J Tatem, Thomas
  Jaenisch, Christopher~JL Murray, Fatima Marinho, Thomas~W Scott, and Simon~I
  Hay.
\newblock Mapping global environmental suitability for {Z}ika virus.
\newblock {\em {eLife}}, 5, 2016.

\bibitem{Doll2006}
Christopher~N.H. Doll, Jan-Peter Muller, and Jeremy~G. Morley.
\newblock Mapping regional economic activity from night-time light satellite
  imagery.
\newblock {\em Ecological Economics}, 57(1):75--92, 2006.

\bibitem{Angiuli2014}
Emanuele Angiuli and Giovanna Trianni.
\newblock Urban mapping in {L}andsat images based on normalized difference
  spectral vector.
\newblock {\em {IEEE} Geoscience and Remote Sensing Letters}, 11(3):661--665,
  2014.

\bibitem{Tran2019}
Annelise Tran, Assane~Gueye Fall, Biram Biteye, Mamadou Ciss, Geoffrey
  Gimonneau, Mathieu Castets, Momar~Talla Seck, and Véronique Chevalier.
\newblock Spatial modeling of mosquito vectors for {Rift Valley} fever virus in
  {Northern Senegal}: Integrating satellite-derived meteorological estimates in
  population dynamics models.
\newblock {\em Remote Sensing}, 11(9), 2019.

\bibitem{worldhealthorganization_2016}
{World Health Organization}.
\newblock Mosquito-borne diseases, 2016.

\bibitem{Scott1997}
Thomas~W. Scott, John~D. Edman, Pattamaporn Kittayapong, Jonathan~F. Day, and
  Amara Naksathit.
\newblock A fitness advantage for aedes aegypti and the viruses it transmits
  when females feed only on human blood.
\newblock {\em The American Journal of Tropical Medicine and Hygiene},
  57(2):235--239, aug 1997.

\bibitem{Scavuzzo2018}
Juan~M. Scavuzzo, Francisco Trucco, Manuel Espinosa, Carolina~B. Tauro, Marcelo
  Abril, Carlos~M. Scavuzzo, and Alejandro~C. Frery.
\newblock Modeling {Dengue} vector population using remotely sensed data and
  machine learning.
\newblock {\em Acta Tropica}, 185:167 -- 175, 2018.

\bibitem{German2018}
A.~German, M.O. Espinosa, M.~Abril, and C.M. Scavuzzo.
\newblock Exploring satellite based temporal forecast modelling of {Aedes}
  aegypti oviposition from an operational perspective.
\newblock {\em Remote Sensing Applications: Society and Environment}, 11:231 --
  240, 2018.

\bibitem{antonio2017}
Fernando~Jose Antonio, Andreia~Silva Itami, Sergio de~Picoli, Jorge
  Juarez~Vieira Teixeira, and Renio dos Santos~Mendes.
\newblock Spatial patterns of {D}engue cases in {B}razil.
\newblock {\em PloS one}, 12(7):e0180715, 2017.

\bibitem{Liu2017}
Zhuanzhuan Liu, Zhenhong Zhang, Zetian Lai, Tengfei Zhou, Zhirong Jia, Jinbao
  Gu, Kun Wu, and Xiao-Guang Chen.
\newblock Temperature increase enhances {A}edes albopictus competence to
  transmit {D}engue virus.
\newblock {\em Frontiers in Microbiology}, 8, 2017.

\bibitem{ritchie1984}
SCOTT~A Ritchie.
\newblock The production of {A}edes aegypti by a weekly ovitrap survey
  [{D}engue, {C}aribbean].
\newblock {\em Mosquito News}, 1984.

\bibitem{eiras2018new}
Alvaro~E Eiras, Marcelo~C Resende, Jos{\'e}~L Acebal, and Kelly~S Paix{\~a}o.
\newblock New cost-benefit of {B}razilian technology for vector surveillance
  using trapping system.
\newblock In {\em From Local to Global Impact of Mosquitoes}. IntechOpen, 2018.

\bibitem{Lary2016}
David~J. Lary, Amir~H. Alavi, Amir~H. Gandomi, and Annette~L. Walker.
\newblock Machine learning in geosciences and remote sensing.
\newblock {\em Geoscience Frontiers}, 7(1):3--10, 2016.

\bibitem{Karpatne2018}
A.~{Karpatne}, I.~{Ebert-Uphoff}, S.~{Ravela}, H.~A. {Babaie}, and V.~{Kumar}.
\newblock Machine learning for the geosciences: Challenges and opportunities.
\newblock {\em IEEE Transactions on Knowledge and Data Engineering},
  31(8):1544--1554, 2019.

\bibitem{Yao2009}
Qing Yao, Zexin Guan, Yingfeng Zhou, Jian Tang, Yang Hu, and Baojun Yang.
\newblock Application of support vector machine for detecting {R}ice diseases
  using shape and color texture features.
\newblock In {\em 2009 International Conference on Engineering Computation}.
  {IEEE}, 2009.

\bibitem{Breiman2001}
Leo Breiman.
\newblock Random forests.
\newblock {\em Machine Learning}, 45(1):5--32, 2001.

\bibitem{Biau2016}
G{\'e}rard Biau and Erwan Scornet.
\newblock A random forest guided tour.
\newblock {\em TEST}, 25(2):197--227, 2016.

\bibitem{McCullagh1989}
P.~McCullagh and J.A. Nelder.
\newblock {\em Generalized Linear Models}.
\newblock Chapman and Hall, 2nd edition, 1989.

\bibitem{Jansen2010}
Cassie~C. Jansen and Nigel~W. Beebe.
\newblock The {D}engue vector {A}edes aegypti: what comes next.
\newblock {\em Microbes and Infection}, 12(4):272--279, 2010.

\bibitem{Santos2017}
Alexandre~Rosa dos Santos, Fel{\'{\i}}cio~Santos de~Oliveira, Aderbal~Gomes
  da~Silva, Jos{\'{e}}~Marinaldo Gleriani, Wantuelfer Gon{\c{c}}alves,
  Giselle~Lemos Moreira, Felipe~Gimenes Silva, Elvis Ricardo~Figueira Branco,
  Marks~Melo Moura, Rosane~Gomes da~Silva, Ronie~Silva Juvanhol,
  Ka{\'{\i}}se~Barbosa de~Souza, Carlos Antonio Alvares~Soares Ribeiro,
  Vagner~Tebaldi de~Queiroz, Adilson~Vidal Costa, Alexandre~Sim{\~{o}}es
  Lorenzon, Getulio~Fonseca Domingues, Gustavo~Eduardo Marcatti, Nero
  Lemos~Martins de~Castro, Rafael~Tassinari Resende, Duberli~Elera Gonzales,
  Lucas~Arthur de~Almeida~Telles, Thaisa~Ribeiro Teixeira, Gleissy Mary Amaral
  Dino~Alves dos Santos, and Pedro Henrique~Santos Mota.
\newblock Spatial and temporal distribution of urban heat islands.
\newblock {\em Science of The Total Environment}, 605-606:946--956, 2017.

\bibitem{costa2018national}
Elis{\^a}ngela Martins da~Silva Costa, Rivaldo Ven{\^a}ncio~da Cunha, and Edgar
  Aparecido~da Costa.
\newblock National dengue control program implementation evaluation in two
  border municipalities in {M}ato {G}rosso do {S}ul {S}tate, {B}razil, 2016.
\newblock {\em Epidemiologia e Servi{\c{c}}os de Sa{\'u}de}, 27(4), 2018.

\bibitem{araujo2015aedes}
Helena Ara{\'u}jo, Danilo Carvalho, Rafaella Ioshino, Andr{\'e} Costa-da Silva,
  and Margareth Capurro.
\newblock Aedes aegypti control strategies in {B}razil: incorporation of new
  technologies to overcome the persistence of dengue epidemics.
\newblock {\em Insects}, 6(2):576--594, 2015.

\bibitem{aboutecovec}
{Ecovec Company}.
\newblock About ecovec.

\bibitem{eiras2009preliminary}
{\'A}lvaro~Eduardo Eiras and Marcelo~Carvalho Resende.
\newblock Preliminary evaluation of the" dengue-mi" technology for aedes
  aegypti monitoring and control.
\newblock {\em Cadernos de Sa{\'u}de P{\'u}blica}, 25:S45--S58, 2009.

\bibitem{Melo2012}
Diogo Portella~Ornelas de~Melo, Luciano~Rios Scherrer, and {\'{A}}lvaro~Eduardo
  Eiras.
\newblock Dengue fever occurrence and vector detection by larval survey,
  ovitrap and {MosquiTRAP}: A space-time clusters analysis.
\newblock {\em {PLoS} {ONE}}, 7(7):e42125, jul 2012.

\bibitem{Justice2002}
C.O Justice, J.R.G Townshend, E.F Vermote, E~Masuoka, R.E Wolfe, N~Saleous, D.P
  Roy, and J.T Morisette.
\newblock An overview of {MODIS} land data processing and product status.
\newblock {\em Remote Sensing of Environment}, 83(1-2):3--15, 2002.

\bibitem{Gao1996}
Bo~cai Gao.
\newblock {NDWI}{\textemdash}a normalized difference water index for remote
  sensing of vegetation liquid water from space.
\newblock {\em Remote Sensing of Environment}, 58(3):257--266, 1996.

\bibitem{Wan2004}
Z.~Wan, Y.~Zhang, Q.~Zhang, and Z.-L. Li.
\newblock Quality assessment and validation of the {MODIS} global land surface
  temperature.
\newblock {\em International Journal of Remote Sensing}, 25(1):261--274, 2004.

\bibitem{skofronick2017global}
Gail Skofronick-Jackson, Walter~A Petersen, Wesley Berg, Chris Kidd, Erich~F
  Stocker, Dalia~B Kirschbaum, Ramesh Kakar, Scott~A Braun, George~J Huffman,
  Toshio Iguchi, et~al.
\newblock The global precipitation measurement ({GPM}) mission for science and
  society.
\newblock {\em Bulletin of the American Meteorological Society},
  98(8):1679--1695, 2017.

\bibitem{Estallo2008}
Elizabet~L. Estallo, Mario~A. Lamfri, Carlos~M. Scavuzzo, Francisco
  F.~Ludue{\~{n}}a Almeida, Mar{\'{\i}}a~V. Introini, Mario Zaidenberg, and
  Walter~R. Almir{\'{o}}n.
\newblock Models for predicting {A}edes aegypti larval indices based on
  satellite images and climatic variables.
\newblock {\em Journal of the American Mosquito Control Association},
  24(3):368--376, 2008.

\bibitem{cheong2013assessing}
Yoon Cheong, Katrin Burkart, Pedro Leit{\~a}o, and Tobia Lakes.
\newblock Assessing weather effects on {D}engue disease in {M}alaysia.
\newblock {\em International Journal of Environmental Research and Public
  Health}, 10(12):6319--6334, 2013.

\bibitem{Bishop1996}
Christopher~M. Bishop.
\newblock {\em Neural Networks for Pattern Recognition}.
\newblock Oxford University Press, 1996.

\bibitem{Sammut2017}
Claude Sammut and Geoffrey~I. Webb.
\newblock {\em Encyclopedia of Machine Learning and Data Mining}.
\newblock Springer, 2017.

\bibitem{Rlanguage}
{R Core Team}.
\newblock {\em R: A Language and Environment for Statistical Computing}.
\newblock R Foundation for Statistical Computing, Vienna, Austria, 2018.

\bibitem{louppe2013understanding}
Gilles Louppe, Louis Wehenkel, Antonio Sutera, and Pierre Geurts.
\newblock Understanding variable importances in forests of randomized trees.
\newblock In {\em Advances in Neural Information Processing Systems}, pages
  431--439, 2013.

\bibitem{Zhang2016}
L.~{Zhang}, L.~{Zhang}, and B.~{Du}.
\newblock Deep learning for remote sensing data: A technical tutorial on the
  state of the art.
\newblock {\em IEEE Geoscience and Remote Sensing Magazine}, 4(2):22--40, 2016.

\bibitem{Lary2010}
David~John Lary.
\newblock {\em Geoscience and Remote Sensing New Achievements}, chapter
  Artificial Intelligence in Geoscience and Remote Sensing, pages 105--128.
\newblock IntechOpen, 2010.

\bibitem{Riedmiller1994}
M.~Riedmiller.
\newblock Rprop - description and implementation details.
\newblock Technical report, University of Karlsruhe, 1994.

\bibitem{Breiman1984}
Leo Breiman, Jerome Friedman, Charles~J. Stone, and R.A. Olshen.
\newblock {\em Classification and Regression Trees}.
\newblock Chapman and Hall/CRC, 1984.

\bibitem{Krzywinski2017}
Martin Krzywinski and Naomi Altman.
\newblock Classification and regression trees.
\newblock {\em Nature Methods}, 14:757--758, 2017.

\bibitem{Hollander2013}
Myles Hollander, Douglas~A. Wolfe, and Eric Chicken.
\newblock {\em Nonparametric Statistical Methods}.
\newblock John Wiley and Sons, third edition, 2013.

\bibitem{chen2010lagged}
Szu-Chieh Chen, Chung-Min Liao, Chia-Pin Chio, Hsiao-Han Chou, Shu-Han You, and
  Yi-Hsien Cheng.
\newblock Lagged temperature effect with mosquito transmission potential
  explains {D}engue variability in southern {T}aiwan: insights from a
  statistical analysis.
\newblock {\em Science of the Total Environment}, 408(19):4069--4075, 2010.

\bibitem{barsante2014model}
LS~Barsante, KS~Paix{\~a}o, KH~Laass, RTN Cardoso, AE~Eiras, and JL~Acebal.
\newblock A model to predict the population size of the dengue fever vector
  based on rainfall data.
\newblock {\em arXiv preprint arXiv:1409.7942}, 2014.

\bibitem{kamimura2002effect}
Kiyoshi Kamumura, Ines~Tomoco Matsuse, Hanako Takahashi, Jun Komukai, Takayo
  Fukuda, Kayo Suzuki, Miho Aratani, Yoshikazu Shirai, and Motoyoshi Mogi.
\newblock Effect of temperature on the development of {A}edes aegypti and
  {A}edes albopictus.
\newblock {\em Medical Entomology and Zoology}, 53(1):53--58, 2002.

\bibitem{Tjaden2013}
Nils~Benjamin Tjaden, Stephanie~Margarete Thomas, Dominik Fischer, and Carl
  Beierkuhnlein.
\newblock Extrinsic incubation period of {D}engue: Knowledge, backlog, and
  applications of temperature dependence.
\newblock {\em {PLoS} Neglected Tropical Diseases}, 7(6):e2207, 2013.

\bibitem{trpivs1972dry}
M~Trpi{\v{s}}.
\newblock Dry season survival of {A}edes aegypti eggs in various breeding sites
  in the {Dar es Salaam} area, {T}anzania.
\newblock {\em Bulletin of the World Health Organization}, 47(3):433, 1972.

\bibitem{fuller2009nino}
Douglas~O Fuller, A~Troyo, and John~C Beier.
\newblock El nino southern oscillation and vegetation dynamics as predictors of
  {D}engue fever cases in {C}osta {R}ica.
\newblock {\em Environmental Research Letters}, 4(1):014011, 2009.

\end{thebibliography}

\end{document}